\def\channel [#1]{h(#1)}
\def\channel{h}
\def\Ar{A_{\rm R}}
\def\FOV{\Theta}
\def\mode{\gamma}
\def\angleLOSandTX{\phi}
\def\rect[#1]{{\Pi}\left(#1\right)}
\def\gain{g}
\def\reflectiveIndex{n}
\def\height{\ell}
\def\distance{d}
\def\distanceTotal{r}

\documentclass[journal]{IEEEtran}
\usepackage{array}

\usepackage[cmex10]{amsmath}
\usepackage{amssymb}
\usepackage{algorithmic}
\usepackage{acronym}
\usepackage{graphicx}
\usepackage{subfig}
\usepackage{float}
\usepackage[noadjust]{cite}
\usepackage{etoolbox}%
\usepackage{amsthm}%
\usepackage{xcolor}
\usepackage{multirow}

\theoremstyle{remark}
\newtheorem{theorem}{Theorem}

\newtheorem{case}{Case}
\newtheorem{remark}{Remark}
\newtheorem{lemma}{Lemma}

\newcolumntype{C}[1]{>{\centering\let\newline\\\arraybackslash\hspace{0pt}}m{#1}}

\begin{document}

\title{Impact of Random Receiver Orientation on \\ Visible Light Communications Channel}

\author{\IEEEauthorblockN{Yusuf Said Ero\u{g}lu, Yavuz Yap{\i}c{\i}, and \.{I}smail G\"{u}ven\c{c}}\\
\IEEEauthorblockA{Department of Electrical and Computer Engineering, North Carolina State University, Raleigh, NC\\
Email: \{yeroglu, yyapici, iguvenc\}@ncsu.edu}}

\author{Yusuf Said Ero\u{g}lu, Yavuz Yap{\i}c{\i}, and \.{I}smail G\"{u}ven\c{c},~\IEEEmembership{Senior~Member,~IEEE}
\thanks{This work is supported in part by NSF CNS awards 1422354 and 1422062.}
\thanks{The authors are with the Department of Electrical and Computer Engineering, North Carolina State University, Raleigh, NC (e-mail:~\{yeroglu, yyapici, iguvenc\}@ncsu.edu).}
}

\maketitle
\begin{abstract}
Visible Light Communications (VLC) has been studied thoroughly in recent years as an alternative or complementary technology to radio frequency communications. The reliability of VLC channels highly depends on the availability and alignment of line of sight links. In this work, we study the effect of random receiver orientation for mobile users over VLC downlink channels, which affects the existence of line of sight links and the receiver field of view. Based on the statistics of vertical receiver orientation and user mobility, we develop a unified analytical framework to characterize the statistical distribution of VLC downlink channels, which is then utilized to obtain the outage probability and the bit error rate. Our analysis is generalized for arbitrary distributions of receiver orientation/location for a single transmitter, and extended to multiple transmitter case for certain scenarios. Extensive Monte Carlo simulations show a perfect match between the analytical and the simulation data in terms of both the statistical channel distribution and the resulting bit error rate. Our results also characterize the channel attenuation due to random receiver orientation/location for various scenarios of interest.  

\end{abstract}

\begin{IEEEkeywords}
Channel statistics, Internet-of-Things (IoT), light-fidelity (Li-Fi), probability density function (pdf), random user orientation, optical wireless communications (OWC), quality of service (QoS).
\end{IEEEkeywords}

\section{Introduction}
Visible light communication (VLC) is an emerging technology that can achieve illumination and communication simultaneously, hence improving energy-efficiency by using existing lighting infrastructure~\cite{Richardson13VLC,Haas14VLCBeyond,Mohapatra15VLCSurvey}. Along with the wide-scale deployment of energy efficient light emitting diodes (LEDs) as the primary luminary, next-generation wireless networks leveraging VLC techniques appear to be even more promising. As recent experiments have revealed, VLC networks can provide data rates as large as multiple Gigabits per second~\cite{Ciaramella2012GbpsVLC, Brien2013GbpsVLC, Haas2014GbpsVLC}, making it a powerful alternative or a complementary technology to conventional radio-frequency (RF) counterparts. 

The propagation through VLC channels can be highly directional \cite{Eroglu_JSAC, eroglu_2015, Alphan_JLT} and communication mainly relies on the availability of line-of-sight (LOS) links. In practice, however, the field-of-view (FOV) of VLC receiver is usually limited, which in turn appears as a barrier in providing seamless network connectivity. The hybrid RF/VLC networks~\cite{Haas2017HybLiFi, Ding2017HybVLC, Wang2017HybVLC, Alouini2016HybVLC, 6011734} and relay-assisted cooperative VLC systems~\cite{Jimenez2015CoOWT, Uysal2015RelAss, Chowdhury2013CooMulCon} are two main research directions to circumvent FOV constraints and extend the network coverage as desired. Furthermore, as the density and mobility of VLC receivers increase along with the use of wearable sensors and Internet-of-Things (IoT) devices~\cite{Elgala2017VLCIoT}, sophisticated dynamics are emerging with FOV constraints and LOS reliability.

The receiver orientation and mobility are two major obstacles affecting the availability of LOS links in VLC networks. Their direct influence on the existence of LOS links and signal quality become even more significant especially when both these features are varying randomly. It is therefore vital to investigate the effect of the receiver orientation and mobility over VLC networks with practical FOV constraints. In~\cite{Uysal2017MobVLC,Ghassemlooy2015DynVLC}, a mobile VLC channel is considered with the goal of characterizing the channel impulse response (CIR) through ray-tracing simulations and laboratory measurements. The user mobility is handled by considering probabilistic noisy and outdated channel state information (CSI) models in~\cite{Karagiannidis2017OnPerVLC}. The ergodic capacity of a mobile VLC scenario is evaluated in~\cite{Sun2017ErgCapVLC} for randomly distributed user locations. Although these recent studies consider the mobility over VLC networks, they all assume fixed and vertically upward receiver orientation without any variation.

The impact of receiver orientation on VLC networks has received very limited attention in the literature. In~\cite{Haas2016ArbRecOri}, a cellular light-fidelity (Li-Fi) network is considered for access point (AP) selection, where the receiver orientation appears to have a significant effect on the user quality of service (QoS) and overall load balancing. The handover mechanism is investigated in \cite{Ghassemlooy2015DynVLC} for mobile Li-Fi networks, and the effect of receiver orientation is evaluated through a geometric approach involving rotation matrix computations. None of these works consider the effect of \textit{random} receiver orientation on VLC channel statistics. The outage performance of an indoor VLC system with random receiver orientation is studied in~\cite{Aveneau2015ImpRecOri} with experimental evaluations. However, this study does not analytically evaluate the effect of random receiver orientation on outage performance. 

In this work, we investigate the effect of the receiver orientation and mobility on the statistics of VLC downlink channels in single and multiple LED scenarios, which has not been studied in the literature before within a broad scope. The contribution of this paper, which is substantially improved version of \cite{Eroglu2017VerOriC}, can be summarized as follows:
\begin{itemize}
\item[i.] We develop a unified analytical framework which derives the statistical distribution of VLC downlink channels explicitly in the presence of random receiver orientation and mobility. The statistical distribution includes the cumulative distribution function (cdf) and the probability density function (pdf) of the channel gain, which enables obtaining the outage probability and the bit error rate (BER), respectively. The channel distribution is characterized in a general form so that any random statistics of the orientation and mobility can be employed directly. The analytical findings are verified through extensive simulation data matching in all cases of interest.
\item[ii.] The nonlinear effect of the receiver FOV is integrated into the analytical framework parametrically, which enables the analysis of channel statistics and error performance for specific FOV chosen from a broad range of values. 
\item[iii.] The proposed framework rigorously handles the single LED and two LEDs cases. In addition, extension of the statistical findings to multiple LED settings are also investigated. The results verify the immediate intuitions that wider FOV and multiple LED deployment can be viable solutions in coping with the adverse effects of random receiver orientation and mobility.
\end{itemize}

The rest of this paper is organized as follows. Section~\ref{sec:system_model} introduces the system model. Section~\ref{sec:singleLED} presents distribution of the square-channel gain for a single LED case, whereas Section~\ref{sec:binaryLED} investigates channel statistics for a specific scenario with two LEDs. Section~\ref{sec:multipleLED} discusses the applicability of the findings for the two LEDs setting to more general multiple LED cases. Finally, Section~\ref{sec:results} presents the respective numerical results, and Section~\ref{sec:conclusion} concludes the paper.

\textit{Notations:} $\mathcal{N}(\mu,\sigma^2)$ denotes the real valued Gaussian distribution with the mean $\mu$ and the variance $\sigma^2$, ${\mathcal{U}[a,b]}$ denotes the continuous uniform distribution over the interval ${[a,b]}$, and $\mathcal{R}(\sigma)$ denotes the Rayleigh distribution with the scale parameter $\sigma$. The trigonometric functions $\cos^{{-}1}(\cdot)$ and $\tan^{{-}1}(\cdot)$ represent the inverse $\cos(\cdot)$ and $\tan(\cdot)$, respectively. $\delta (a)$ is the Dirac delta function taking $1$ if $a\,{=}\,0$, and $0$ otherwise.

\begin{figure}[t]
\centering
\includegraphics[width=3.4in]{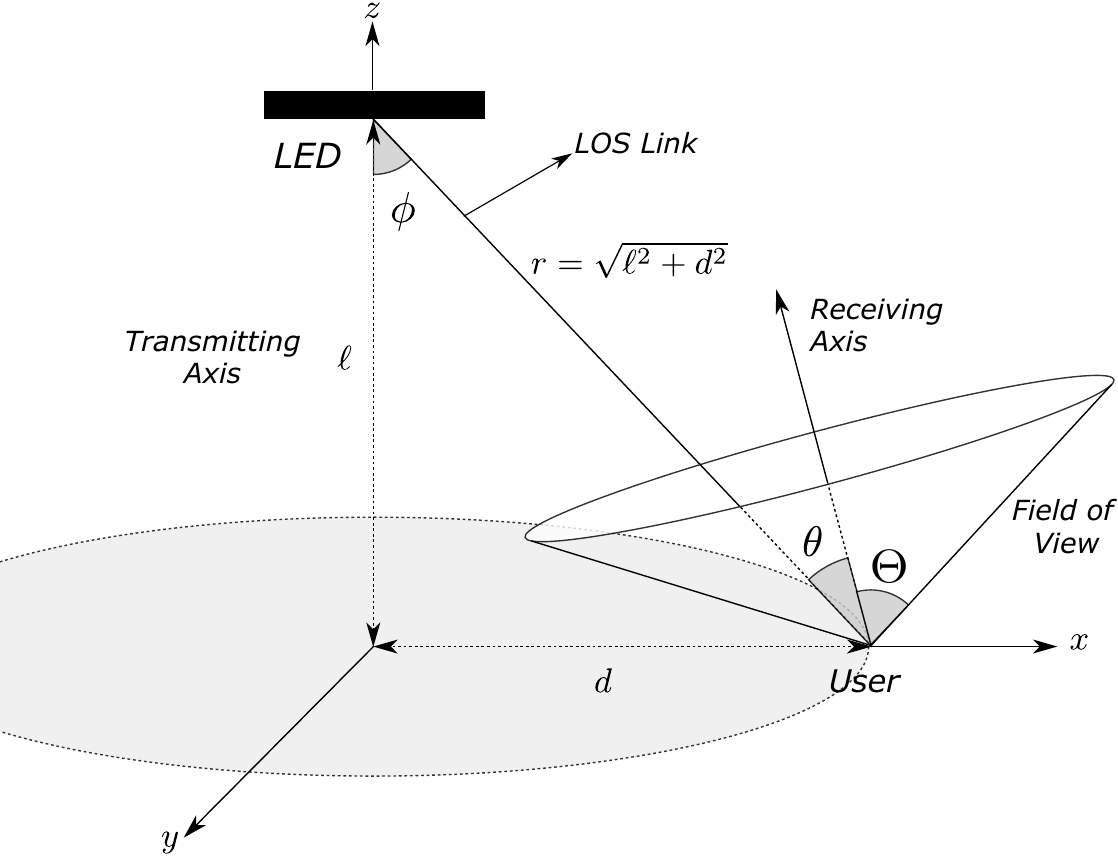}
\caption{VLC downlink transmission model with random receiver orientation $\theta$.}
\label{fig:setup}
\end{figure}

\section{System Model}\label{sec:system_model}
We consider an indoor VLC downlink transmission scenario with multiple LEDs, where a single user is served by a single LED at a given time. The interaction between an LED and the user over a point-to-point LOS link is sketched in Fig.~\ref{fig:setup}. The corresponding direct current (DC) channel gain is given as~\cite{219552}
\begin{align}
\channel =\frac{(\mode+1)\Ar \gain}{2\pi\distanceTotal^2}\cos^\mode(\angleLOSandTX) \cos(\theta)\rect[ \frac{\theta}{\FOV} ],
\label{eq:LOSchannel}
\end{align}
where $\distanceTotal$ is the LOS distance between the LED and the user, $\angleLOSandTX$ is the angle of irradiance, $\theta$ is the angle of incidence, $m\,{=}\,{-}1/\log_2(\cos(\Phi_{1/2}))$ is the Lambertian order with $\Phi_{1/2}$ being the half-power beamwidth of the LED, $\Ar$ is the detection area of the receiver, $\gain$ is the gain of the optical concentrator given by $\reflectiveIndex^2 / \sin^2(\FOV)$ with $\reflectiveIndex$ being the reflective index, and $\FOV$ is the FOV angle of the receiver. The notation $\rect[x]$ represents a rectangular function given as
\begin{align}
\rect[x]\triangleq\begin{cases}
1 & \mbox{for  } |x| \leq 1  \\
0 & \mbox{for  } |x| > 1
\end{cases}~,
\end{align} 
and hence, $\rect[\theta/\FOV]$ implies that the channel gain is zero if $\theta$ is larger than $\FOV$, or equivalently the LED is outside the receiver FOV. The observation model for the point-to-point transmission scenario at $k$th discrete time instant is given as
\begin{align}
y_k = h \, a_k + v_k ,
\end{align}
where $y_k$ is the received signal, $a_k$ is the transmitted symbol chosen from a modulation alphabet $\mathcal{A}$ with the average \emph{transmit energy} $E_s$, 
and $v_k$ is the white Gaussian noise with zero mean and variance $N_0/2$. Without loss of generality, assuming the binary on-off-keying (OOK) modulation, the average probability of bit error is given as
\begin{align}\label{eqn:avPe}
P_{\rm e} = \int_{0}^{\infty} Q\left(\sqrt{\left( E_{\rm s}/N_0 \right) \varphi} \right) f_{\channel^2}(\varphi) \, \rm d\varphi,
\end{align}
where $f_{\channel^2}(\varphi)$ is the pdf of the square-channel expression denoted as $\channel^2$, and $Q(\cdot)$ is the $Q$-function~\cite{ProakisDigiComm}. In order to calculate the BER in \eqref{eqn:avPe}, the pdf of $\channel^2$ needs to be known. 

In the sequel, we will characterize the point-to-point VLC channel in \eqref{eq:LOSchannel} when the user orientation fluctuates randomly around the vertical axis, which is represented by the random incidence angle $\theta$. Assuming a stochastic distribution for this fluctuation in the vertical direction, we will derive the distribution of the square-channel $\channel^2$, and evaluate the impact of this random behavior on the BER statistics via \eqref{eqn:avPe}. To this end, we will consider single and multiple LED scenarios with the deterministic and random user deployment cases, separately, in the subsequent sections. 

\section{Square-Channel Distribution for Single LED}\label{sec:singleLED}
In this section we derive the channel statistics for a single LED scenario as shown in Fig.~\ref{fig:setup}. We can rearrange \eqref{eq:LOSchannel} as follows: 
\begin{align}\label{eq:chan_simplified}
h = \frac{(\mode+1)\Ar\ell^m\gain}{2\pi}\left(\ell^2 + d^2\right)^{-\frac{\mode+2}{2}}\cos(\theta) \, \Pi \left( \frac{\theta}{\FOV} \right) ,
\end{align}
where we employ the geometrical relations $r\,{=}\,\sqrt{\distance^2\,{+}\,\height^2}$ and $\cos(\phi)\,{=}\,\height/\sqrt{\distance^2\,{+}\,\height^2}$ from Fig.~\ref{fig:setup} with $\height$ and $\distance$ being the vertical and horizontal distances between the LED and the user, respectively. 

We assume two different scenarios regarding the FOV effect of the receiver while analyzing the statistical behavior of the square-channel under receiver orientation fluctuations. In the first scenario, we assume that the FOV of the receiver is wide enough (characterized by a large $\FOV$), therefore the LED is always within the FOV. This is a simplistic scenario that enables the derivation of square-channel statistics without any nonlinear effects arising from FOV restrictions~\cite{Uysal15ChanModel}, and referred to as ``\textit{wide FOV}'' scenario in this paper. The second scenario assumes a more general setting by assuming ``\textit{narrow FOV}'', where the LED might be either inside or outside the FOV depending on the vertical user orientation and specifics of the geometry in Fig.~\ref{fig:setup}. This second scenario considers all possible geometrical interactions between the LED and user, but complicates the derivation of the desired channel statistics.

\subsection{Deterministic User Location and Wide FOV} \label{sec:wideFOV_single}
We first assume wide FOV where the incidence angle $\theta$ in \eqref{eq:chan_simplified} is always smaller than $\FOV$, which implies $\rect[\theta/\FOV]\,{=}\,1$. In addition, the user location is assumed to be chosen in a deterministic fashion such that the horizontal distance $d$ is a nonrandom variable. Then, the random part of the channel gain in \eqref{eq:chan_simplified} is $h_\theta = \cos(\theta)$. The distribution of the square-channel $h^2$ can be derived by considering the cdf of $h_\theta^2\,{=}\,\cos^2(\theta)$ given as
\begin{align}
F_{h^2_\theta}(x) &= \Pr \left\lbrace \cos^2(\theta) < x \right\rbrace , \label{eqn:cdf_fixloc_wide_1} \\
&= \Pr \left\lbrace \theta > \frac{1}{2}\cos^{-1}(2x-1) \right\rbrace . \label{eqn:cdf_fixloc_wide_2}
\end{align}
Note that, the probability in \eqref{eqn:cdf_fixloc_wide_1} is always $1$ for $x\,{\geq}1\,$, and $0$ for $x\,{<}\,0$, and we therefore limit $x$ to the interval $[0,1]$ while analyzing \eqref{eqn:cdf_fixloc_wide_2}. Defining $F_\theta(.)$ to be the cdf of the random incidence angle $\theta$, the cdf of the $h_\theta^2$ is obtained by rearranging \eqref{eqn:cdf_fixloc_wide_2} as follows
\begin{align}
F_{h^2_\theta}(x) = 1 - F_\theta \left(\frac{1}{2}\cos^{-1}(2x-1) \right). \label{eqn:cdf_fixloc_wide_3} 
\end{align}
The corresponding pdf can be computed by taking derivative of \eqref{eqn:cdf_fixloc_wide_3} with respect to $x$, and is given as
\begin{align}
f_{h^2_\theta}(x) &= \frac{c_{\rm \theta}}{\sqrt[]{4x(1-x)}}f_{\theta}\left(\frac{1}{2}\cos^{-1}(2x-1)\right), \label{eqn:pdf_fixloc_wide}
\end{align}
for $0 \leq x \leq 1$, and $0$ otherwise. In \eqref{eqn:pdf_fixloc_wide}, $c_{\rm \theta}$ is the normalization constant, and $f_{\theta}(.)$ is the pdf of the random angle $\theta$. Denoting the deterministic part of \eqref{eq:chan_simplified} as $h_c$ such that $h\,{=}\,h_c h_\theta$, the cdf and pdf of the square-channel is readily given as 
\begin{align}
F_{h^2}(x) = F_{h^2_\theta} \left(\frac{x}{h_c^2}\right) \,, \quad f_{h^2}(x) = \frac{1}{h_c^2} f_{h^2_\theta} \left(\frac{x}{h_c^2}\right) \,, \label{eqn:dist_hSq}
\end{align}
which are defined in the most general form such that any distribution for the random angle $\theta$ can be used directly via \eqref{eqn:cdf_fixloc_wide_3}-\eqref{eqn:pdf_fixloc_wide}.

\subsection{Deterministic User Location with Narrow FOV} \label{sec:narrowFOV_single}
When we assume a narrow FOV, the point-to-point LOS link in Fig.~\ref{fig:setup} can be outside the receiver FOV because of the random orientation of the user around the vertical axis. In this case the square-channel $h^2$ can be derived by considering the associated random part $h_\theta^2\,{=}\,\cos^2(\theta)\,\rect[\theta/\FOV]$, with the cdf given as
\begin{align}
F_{h^2_\theta}(x) &= \Pr \left\lbrace \cos^2(\theta)\rect[\theta/\FOV] < x \right\rbrace \nonumber \\
&= \Pr \left\lbrace \cos^2(\theta) \,{<}\, x, 0 \,{\leq}\, \theta {\leq}\, \FOV \right\rbrace  + \Pr\left\lbrace x {>} 0, \FOV \,{<}\, \theta \right\rbrace , \label{eqn:cdf_fixloc_nar_1} 
\end{align}
where the first and second probabilities in \eqref{eqn:cdf_fixloc_nar_1} represent the cases where the LOS link is within the FOV and outside the FOV, respectively. 

For ease of representation, we define the function $\bigtriangleup_\theta(a,b)$, which represents the probability of the random variable $\theta$ being in an interval $(a,b\,]$ with arbitrary real-valued variables $a,b \,{\in}\, \mathbb{R}$, and is given as follows
\begin{align}\label{eqn:delta_theta}
\bigtriangleup_\theta(a, b) = \Pr\{a<\theta\leq b \} = \begin{cases}
F_\theta(b) - F_\theta(a) \!\!\!&\mbox{for } a \leq b \\
0 &\mbox{for } a > b
\end{cases}.
\end{align}
Then, following the strategy in obtaining \eqref{eqn:cdf_fixloc_wide_2}, the cdf in \eqref{eqn:cdf_fixloc_nar_1} becomes
\begin{align}
\!\!\!\!F_{h^2_\theta}(x) &= \Pr\left\lbrace \frac{1}{2}\cos^{-1}(2x{-}1) \,{<} \, \theta\,{\leq}\,\FOV \right\rbrace {+} \Pr\{ \theta \,{>}\, \FOV \} \,, \label{eqn:cdf_fixloc_nar_2} \\
&= \bigtriangleup_\theta\left(\frac{1}{2}\cos^{{-}1}(2x-1), \FOV\right) + 1-F_{\theta}(\FOV), \label{eqn:cdf_fixloc_nar_3} 
\end{align}
for $0 \,{\leq}\, x \,{\leq}\, 1$, equal to $0$ for $x \,{<}\, 0$, and $1$ for $x \,{>}\, 1$. Note that, when the FOV takes a large value, the random angle $\theta$ is always less than $\FOV$ implying $\Pr\{ \theta \,{>}\, \FOV \}\,{=}\,0$ and $F_\theta(\FOV)\,{=}\,1$, and \eqref{eqn:cdf_fixloc_nar_2} readily yields the cdf expression in \eqref{eqn:cdf_fixloc_wide_3} of the wide FOV scenario, as expected. 

We observe that the first term involving the function $\bigtriangleup_\theta(\cdot)$ is zero over the interval $0 \,{\leq}\, x \,{<}\, \cos^2(\FOV)$ by the definition in \eqref{eqn:delta_theta}, and the cdf in \eqref{eqn:cdf_fixloc_nar_3} becomes equal to $1-F_{\theta}(\FOV)$ over this interval. The cdf is $0$ for $x\,{<}\,0$, which means the function in \eqref{eqn:cdf_fixloc_nar_3} is discontinuous at $x\,{=}\,0$. With this observation, we can give the corresponding pdf as follows  
\begin{align}
f_{h^2_\theta}(x) &= c_{\rm \theta} \frac{\partial}{\partial x} \bigtriangleup_\theta\left(\frac{1}{2}\cos^{{-}1}(2x{-}1), \FOV\right) \nonumber\\ &  \qquad \qquad  \qquad  \qquad \qquad {+}\, (1\,{-}\,F_{\theta}(\FOV) ) \, \delta(x),
\label{eqn:pdf_fixloc_nar}
\end{align}
for $0 \leq x \leq 1$, and $0$ otherwise. In \eqref{eqn:pdf_fixloc_nar}, $c_{\rm \theta}$ is the normalization constant\footnote{Since the pdf has a Dirac delta function term of size $1-F_\theta(\FOV)$, the normalization constant normalizes the integral sum of the other term to $F_\theta(\FOV)$ so that overall integral sum is equal to one.}, and the Dirac delta function $\delta(x)$ appears as a result of the discontinuity of the cdf in \eqref{eqn:cdf_fixloc_nar_3} at $x\,{=}\,0$. The partial derivative in \eqref{eqn:pdf_fixloc_nar} is given as
\begin{align}
\frac{\partial}{\partial x} & \bigtriangleup_\theta\left(\frac{1}{2}\cos^{{-}1}(2x{-}1), \FOV\right) = \nonumber \\
& \qquad \qquad \frac{1}{\sqrt[]{4x(1-x)}}f_{\theta}\left(\frac{1}{2}\cos^{-1}(2x-1)\right), \label{eqn:pdf_delta_fixloc_nar}
\end{align}
for $\cos^{2}(\FOV) \leq x <1$, and $0$ otherwise. Finally, the desired cdf and pdf of the square-channel $h^2$ can be obtained readily by \eqref{eqn:dist_hSq}.  

\begin{figure}[t]
\centering
\includegraphics[width=3.4in]{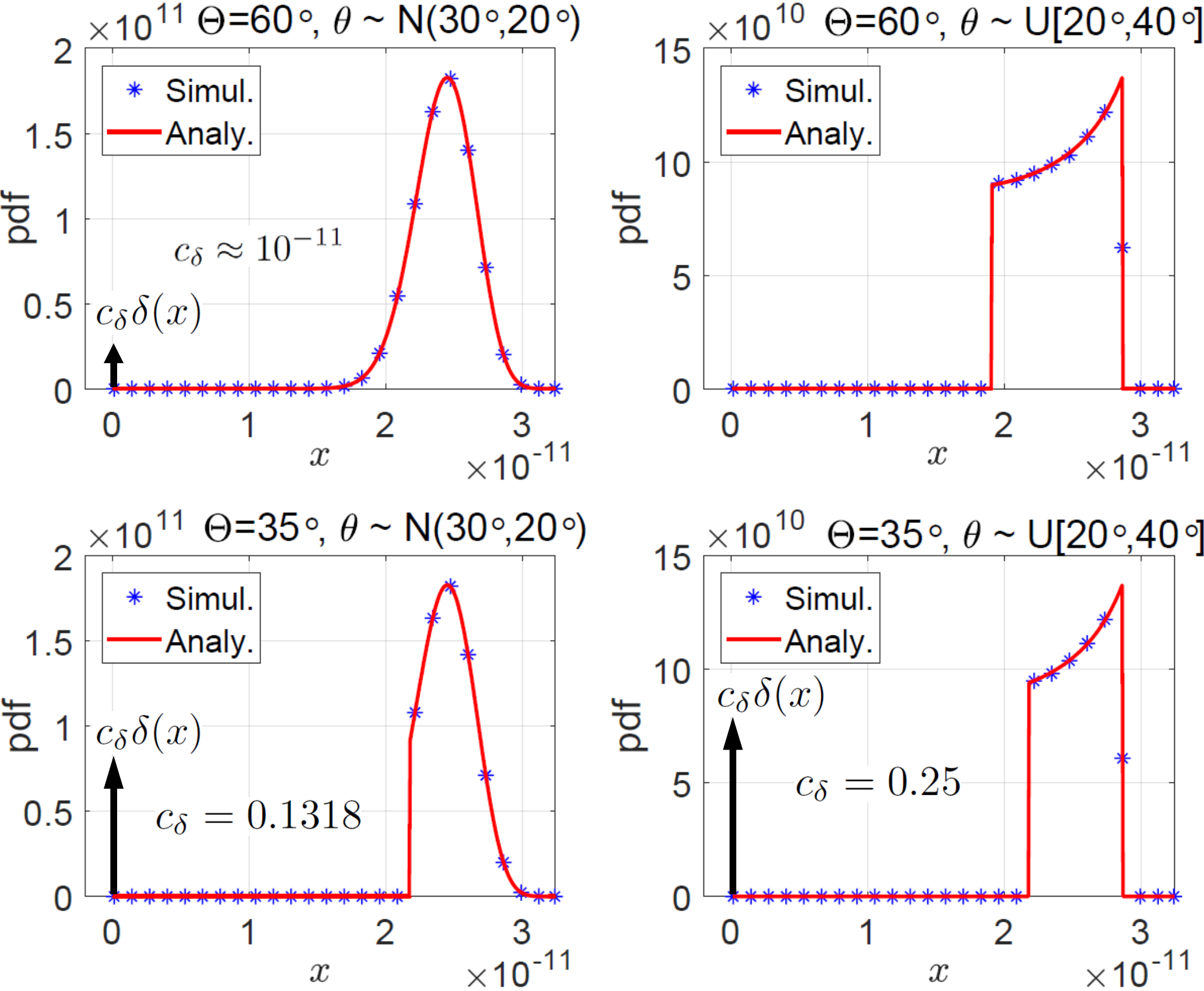}
\caption{The pdf of square-channel for different receiver orientation distributions and FOV angles. Parameters are $d\,{=}\,2.5$~m, $\ell\,{=}\,3$~m, $\Phi_{1/2}\,{=}\,60^{\circ}$, $g\,=\,1$, and $\Ar\,{=}\,1$~cm$^2$.}
\label{fig:FixedLocPDF}
\end{figure}

Note that, the impact of the narrow FOV on the square-channel pdf can be interpreted by comparing \eqref{eqn:pdf_fixloc_wide} and \eqref{eqn:pdf_fixloc_nar} with the help of \eqref{eqn:pdf_delta_fixloc_nar}. We observe that limiting the receiver FOV with a narrow angle $\FOV$ introduces a Dirac delta weighted by $1\,{-}\,F_{\theta}(\FOV)$ and discards the pdf portion within the interval of $0 \leq x < \cos^{2}(\FOV)$, which in turn causes the magnitude of the pdf to be weighted over the interval $\cos^{2}(\FOV) \leq x <1$. To illustrate this issue, we have provided square-channel pdfs for different receiver orientation distributions in  Fig.~\ref{fig:FixedLocPDF} for two different FOV angles. In the top figures, pdfs for $\FOV~=~60^\circ$ are given. The delta function does not occur for uniform distribution, and it is arbitrarily small for normal distribution. It is because the FOV is large, and probability of leaving LED out of FOV ($\Pr\{\theta>\FOV\}$) is zero or arbitrarily small. In the bottom figures, pdfs for $\FOV~=~35^\circ$ are given. We observe that pdf shapes are clipped from left side (from $x = \cos^{2}(\FOV)$), and the area under the clipped shape is accumulated at $x = 0$ which appears as delta function. The smaller values of x corresponds to the larger values of $\theta$, and when $\theta$ is larger than $\FOV$ the channel gain is equal to zero.

\subsection{Random User Location with Wide and Narrow FOVs}\label{sec:random_loc_single_led}
In this section, we consider a mobile user over the $xy$-plane as in Fig.~\ref{fig:setup}, and the associated mobility is captured by choosing the horizontal distance $d$ at random, which corresponds to the random user deployment strategy. This random effect in channel due to the user mobility is represented by $h_d\,{=}\,\left(\ell^2\,{+}\,d^2\right)^{{-}(\mode{+}2)/2}$, and the channel in \eqref{eq:chan_simplified} accordingly becomes $h\,{=}\,h_c\,h_d\,h_\theta$ where $h_\theta\,{=}\,\cos(\theta)\,\rect[\theta/\FOV]$ represents the random part due to the vertical orientation including the effect of the narrow FOV, and $h_c$ is the remaining deterministic part, as before. 

The distribution of the square-channel can now be computed by exploiting the independence of $h_\theta$ and $h_d$, and by employing the property on the distribution of the product of independent random variables as follows~\cite{Springer66Pro}
\begin{align} \label{eqn:cdf_joint} 
F_{h^2}(x) = 1\,{-}\,F_{\theta}(\FOV) + \frac{1}{h_c^2}\int_{\mathcal{R}_y} \frac{1}{y} f_{h^2_d}(y)F_{h^2_\theta}\left(\frac{x}{h_c^2\,y}\right) {\rm d}y \,,
\end{align}
\vspace{-1.5mm}
\begin{align} \label{eqn:pdf_joint} 
f_{h^2}(x) =
\begin{cases}
\frac{c_h}{h_c^2}\int_{\mathcal{R}_y} \frac{1}{y} f_{h^2_d}(y)f_{h^2_\theta}\left(\frac{x}{h_c^2\,y}\right) {\rm d}y & {\mbox for~} 0<x\leq 1 \\
1\,{-}\,F_{\theta}(\FOV) & {\mbox for~} x = 0
\end{cases},
\end{align}
where $\mathcal{R}_y$ is the set of $y$ values for which the function being integrated takes nonzero value, and $c_h$ is the normalization constant. Defining $F_d(.)$ to be the cdf of the random distance $d$, the desired pdf of $h^2_d$ can be found by considering the cdf given as follows
\begin{align}
F_{h^2_d}(y) &= \Pr \left\lbrace (d^2 + \ell^2)^{-(\mode+2)} < y \right\rbrace \nonumber\\
&=\Pr\left\lbrace \distance^2 > y^{-\frac{1}{\mode + 2}} - \height^2 \right\rbrace \nonumber\\
&=1 - F_d \left( \left[ y^{-\frac{1}{\mode + 2}} - \height^2\right]^{1/2}\right), \label{eqn:cdf_location} 
\end{align}
for $0\,{\leq}\,y\,{\leq}\,\height^{{-}2(\mode{+}2)}$, and $1$ for $y\,{>}\,\height^{{-}2(\mode{+}2)}$. Taking the derivative of \eqref{eqn:cdf_location}, the desired pdf of $h^2_d$ is then found to be
\begin{align}
f_{h^2_d}(y) &= c_d \, y^{{-}\frac{\mode{+}3}{\mode{+}2}}\left[ y^{{-}\frac{1}{\mode{+}2}}\,{-}\,\height^2 \right]^{-\frac{1}{2}} \!\! f_d \left( \left[ y^{{-}\frac{1}{\mode{+}2}}\,{-}\,\height^2 \right]^{\frac{1}{2}} \right), \label{eqn:pdf_location}
\end{align}
for $0\,{\leq}\,y\,{\leq}\,\height^{{-}2(\mode{+}2)}$, and $0$ otherwise. In \eqref{eqn:pdf_location}, $c_d$ is the normalization constant, and $f_d(\cdot)$ is the pdf of the random distance $d$. Incorporating \eqref{eqn:cdf_fixloc_nar_3} and \eqref{eqn:pdf_location} into \eqref{eqn:cdf_joint}, we can obtain the cdf of the square-channel $h^2$, which can be applied to both the wide and the narrow FOV settings. Likewise, incorporating \eqref{eqn:pdf_fixloc_nar} and \eqref{eqn:pdf_location} into \eqref{eqn:pdf_joint}, we can obtain the pdf of the square-channel $h^2$. 
As before, the distribution in \eqref{eqn:pdf_joint} is defined in general form such that any distribution for the random incidence angle $\theta$ and the horizontal distance $d$ can be incorporated directly.   

\section{Square-Channel Distribution for Two LEDs} \label{sec:binaryLED}
In this section, we discuss the distribution of the square-channel for a specific two LED transmitters scenario, where the user location varies between two points which are beneath the LEDs, and user orientation is random around the vertical axis. While this scenario does not cover all possible two LED geometries, it might be useful for evaluating the channel performance when a person holding a VLC receiver device is walking in a corridor equipped with VLC transmitters.

\begin{figure}[t]
\centering
\includegraphics[width=3in]{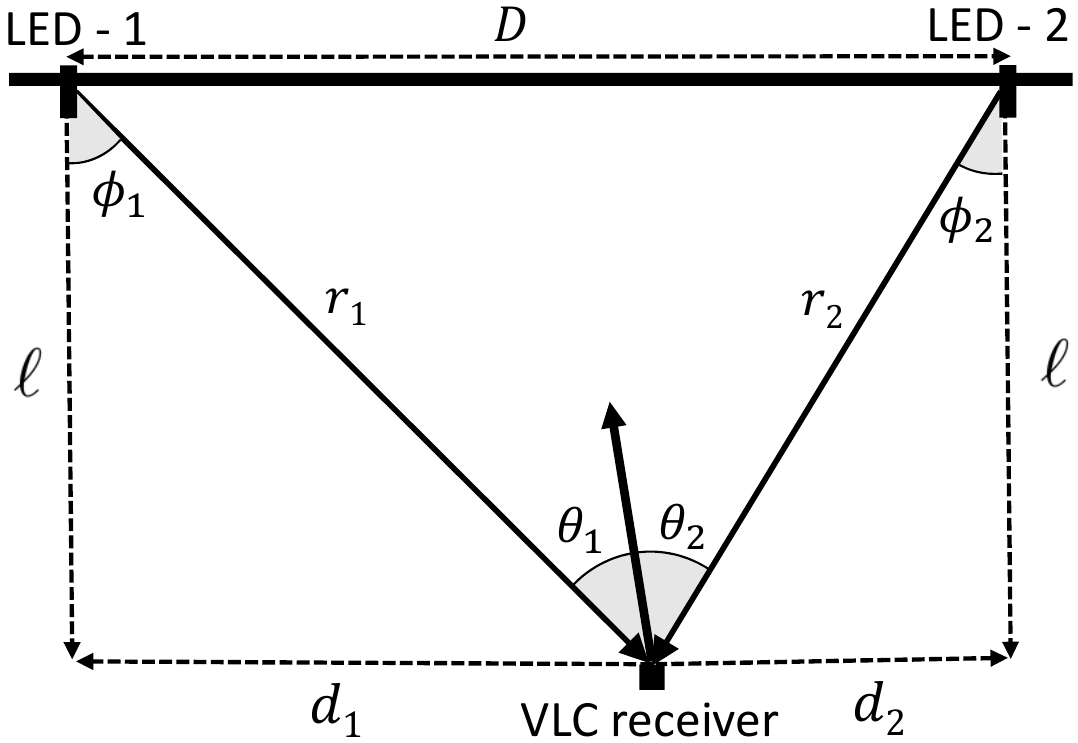}
\caption{VLC receiver with random orientation and two LEDs serving to that user.}
\label{fig:2LED}
\end{figure}

\subsection{Two LEDs Scenario}
In this scenario two LEDs are available to serve a single user as in Fig.~\ref{fig:2LED}, and the user is assigned to the LED with the strongest signal. The resulting instantaneous effective channel is given as
\begin{align}
h_{\rm eff} = \max \left\lbrace h_1^2, h_2^2 \right\rbrace,
\end{align}
where $h_1$ and $h_2$ are the point-to-point channel gains from the first and the second LEDs, respectively, to the user. The channel gains are jointly represented as $h_i\,{=}\,h_c h_{d_i} h_{\theta_i}$ for $i\,{=}\,1,2$, where $h_c\,{=}\, (\mode+1)\Ar\ell^m\gain / 2\pi$ is the constant multiplier not depending on either the location or the orientation of the receiver. Other multipliers are
\begin{align}
h_{d_i} &= (\ell^2 + d_i^2)^{-\frac{\mode+2}{2}}, \label{eqn:h_d_i}\\
h_{\theta_i} &= \cos(\theta_i) \, \Pi\left( \frac{\theta_i}{\FOV}\right), \label{eqn:h_theta_i} 
\end{align}
where $\theta_i$ and $d_i$ are the incidence angle and the horizontal distance of the user with respect to the $i$th LED, respectively, as shown in Fig.~\ref{fig:2LED}. Without loss of generality, we define the random angles $\theta_1$ and $\theta_2$ in the clockwise and counter-clockwise directions, respectively. 

The actual degrees of freedom for both $\{\theta_1,\theta_2\}$ and $\{d_1,d_2\}$ are $1$, so that we have only one independent random variable from each set. We therefore choose $d_1\,{=}\,d$ and $\theta_1\,{=}\,\theta$ as the independent random variables of interest, and $d_2\,{=}\,D-d$ and $\theta_2\,{=}\,\Phi\,{-}\,\theta$ accordingly become the dependent random variables, where $D$ is the distance between the LEDs, and the non-negative angle $\Phi$ is defined geometrically by Fig.~\ref{fig:2LED} as follows
\begin{align}\label{eqn:theta_sum}
\Phi\,{=}\,\phi_1\,{+}\,\phi_2\,{=}\,\tan^{{-}1}(d/\ell)\,{+}\,\tan^{{-}1}((D\,{-}\,d)/\ell).
\end{align}
Note that $\theta_1$ and $\theta_2$ can take negative values, but both of them cannot be negative at the same time, and their sum is equal to $\Phi$ in any case. 

\subsection{Two LEDs with Fixed User Location}
When the user location is not considered to be random, the horizontal distance $d$ turns out to be a deterministic variable. In this case, the cdf of the square-effective channel gain is then given as
\begin{align}
F_{h_{\rm eff}^2}(x) &= \Pr \left\lbrace \max \left\lbrace h_1^2, h_2^2 \right\rbrace < x \right\rbrace \nonumber\\
&= \Pr \left\lbrace h_1^2 < x , h_2^2 < x \right\rbrace , \label{eqn:cdf_nonrand_twoled_1}
\end{align}
where \eqref{eqn:cdf_nonrand_twoled_1} directly follows from~\cite{Papoulis2002Prob}. As the horizontal distance $d$ is not random, we define a new variable $c_i\,{=}\, h_c^2 h_{d_i}^2$ for $i\,{=}\,1,2$, which captures the deterministic feature of the square-effective channel, and is always non-negative by definition. Then, the cdf in \eqref{eqn:cdf_nonrand_twoled_1} becomes
\begin{align}\label{eqn:cdf_nonrand_twoled_2}
F_{h_{\rm eff}^2}(x) &= \Pr \left\lbrace h_{\theta_1}^2 < \frac{x}{c_1}, h_{\theta_2}^2 < \frac{x}{c_2} \right\rbrace,
\end{align}
and employing \eqref{eqn:h_theta_i} yields 
\begin{align}
& F_{h_{\rm eff}^2}(x) = \begin{cases}
\displaystyle \Pr\left\lbrace \mathcal{E}_1,\mathcal{E}_2 \right\rbrace,\!\!\!\! &\mbox{for } |\theta_1| \leq \FOV, |\theta_2| \leq \FOV \\
\displaystyle \Pr\left\lbrace \mathcal{E}_1 \right\rbrace, &\mbox{for } |\theta_1| \leq \FOV, |\theta_2| > \FOV \\
\displaystyle \Pr\left\lbrace \mathcal{E}_2 \right\rbrace, &\mbox{for } |\theta_1| > \FOV, |\theta_2| \leq \FOV \\
1, &\mbox{for } |\theta_1| > \FOV, |\theta_2| > \FOV
\end{cases} \label{eqn:cdf_nonrand_twoled_3},
\end{align}
where the event $\mathcal{E}_i$ is defined for $i\,{=}\,1,2$ as follows
\begin{align}\label{eqn:event}
\mathcal{E}_i : \left\lbrace \theta_i \in \Omega_{\theta_i} \,|\, cos^2\theta_i < \frac{x}{c_i} \right\rbrace \,,
\end{align}
with $\Omega_{\theta_i}$ being the sample space of $\theta_i$. Note that $\mathcal{E}_i$ happens with probability $1$ whenever $x\,{\geq}\,c_i$, and probability $0$ whenever $x\,{<}\,0$ (does not happen at all). We therefore safely assume $x\,{=}\,c_i$ for $x\,{\geq}\,c_i$ and $x\,{=}\,0$ for $x\,{<}\,0$, to keep the argument of the inverse cosine function within the definition interval. 

Before further elaborating the cdf in \eqref{eqn:cdf_nonrand_twoled_3}, we define a new function in the following Lemma, which is actually an extension of \eqref{eqn:delta_theta}.
\begin{lemma} \label{lem:delta_two_int}
Let $\bigtriangledown_\theta(a, b, c, d)$ be a function defined as
\begin{align}\label{eqn:delta_two_int_1}
\bigtriangledown_\theta(a, b, c, d) &= \Pr\{a<\theta\leq b,~ c < \theta \leq d \},
\end{align}
which represents the probability of the random variable $\theta$ being in the intervals $(a,b\,]$ and $(c,d\,]$ jointly, with arbitrary real-valued variables $a,b,c,d \,{\in}\, \mathbb{R}$. Then, \eqref{eqn:delta_two_int_1} can be computed in terms of the cdf of the random variable $\theta$ as follows
\begin{align}\label{eqn:delta_two_int_2}
\bigtriangledown_\theta(a, b, c, d) &= \begin{cases}
F_\theta(b) - F_\theta(a) &\mbox{for } c \leq a,~  d > b \\
F_\theta(d) - F_\theta(c) &\mbox{for } c > a ,~ d \leq b \\
F_\theta(d) - F_\theta(a) &\mbox{for } c \leq a, ~ d \leq b \\
F_\theta(b) - F_\theta(c) &\mbox{for }  c > a ,~ d > b \\
0 & \mbox{otherwise}
\end{cases}.
\end{align}
\end{lemma}
\begin{IEEEproof}
See Appendix~\ref{app:delta_two_int}.
\end{IEEEproof}

Defining $z_i(x) = \frac{1}{2}\cos^{-1}(2\frac{x}{c_i}-1)$ for $i\,{=}\,1,2$, the desired cdf and pdf expressions are given in the next Theorem.
\begin{theorem} \label{the:cdf_nonrand_twoled}
The cdf of the square of the effective channel given in \eqref{eqn:delta_two_int_2} can be expressed as follows
\begin{align}\label{eqn:cdf_nonrand_twoled_4}
F_{h_{\rm eff}^2}(x) = P_1(x) + P_2(x) + P_3(x) + P_4,
\end{align}
where
\begin{align}
P_1(x) &= \bigtriangledown_\theta({-}\FOV, {-}z_1(x), \Phi \,{-}\, \FOV, \Phi\,{-}\,z_2(x)) \nonumber \\
&+ \bigtriangledown_\theta(z_1(x), \FOV, \max(0, \Phi {-} \FOV), \Phi {-} z_2(x)) \nonumber \\
&+ \bigtriangledown_\theta(z_1(x), \FOV, \Phi \,{+}\, z_2(x), \Phi \,{+}\, \FOV), \label{eqn:p1} \\
P_2(x) &= \bigtriangleup_\theta({-}\FOV, \min({-}z_1(x), \Phi {-} \FOV)) \nonumber \\
&+ \bigtriangleup_\theta(z_1(x), \min(\FOV, \Phi \,{-}\, \FOV )), \label{eqn:p2} \\
P_3(x) &= \bigtriangledown_\theta(\Phi \,{-}\, \FOV, \Phi \,{-}\, z_2(x), \FOV, \Phi) \nonumber\\
&+ \bigtriangleup_\theta(\max(\Phi \,{+}\, z_2(x), \FOV), \Phi \,{+}\, \FOV), \label{eqn:p3} \\
P_4 &= \bigtriangleup_\theta(\FOV, \Phi \,{-}\, \FOV) \,{+}\, F_\theta({-}\FOV) \,{+}\, 1 \,{-}\, F_\theta(\Phi \,{+}\, \FOV). \label{eqn:p4}
\end{align}
Furthermore, taking derivative of \eqref{eqn:cdf_nonrand_twoled_4} yields the desired pdf as follows
\begin{align}\label{eqn:pdf_nonrand_twoled}
f_{h_{\rm eff}^2}(x) = c_\theta \displaystyle\sum_{j=1}^{3} \frac{\partial P_j(x)}{\partial \theta} + P_4\delta(x)\,,
\end{align}
The individual derivatives $\frac{\partial P_j(x)}{\partial \theta}$'s in \eqref{eqn:pdf_nonrand_twoled} can be computed using the following derivative expression 
\begin{align}\label{eqn:cdf_theta_del}
\frac{\partial }{\partial x} F_\theta \left(u \,{+}\, v \, z_i(x)\right) = \frac{{-}v}{\sqrt[]{4x \left( c_i\,{-}\,x \right)} }f_\theta\left( u \,{+}\, v \, z_i(x) \right), 
\end{align}
where the argument $u \,{+}\, v \, z_i(x)$ of the cdf function $F_\theta(\cdot)$ is the linear transformation of $z_i(x)$ with the arbitrary transformation variables $u$ and $v$. 
\end{theorem}
\begin{IEEEproof}
See Appendix~\ref{app:nonrand_twoled}.
\end{IEEEproof}

Note that the cdf expression in \eqref{eqn:cdf_nonrand_twoled_4} handles all possible combinations of the incidence angles $\theta_i$'s and the receiver FOV $\FOV$. The $P_j(x)$ functions given in \eqref{eqn:p1}-\eqref{eqn:p3} correspond to the first three cases of \eqref{eqn:cdf_nonrand_twoled_3}, in the same order. The $P_4$ expression given in \eqref{eqn:p4} corresponds to fourth case of \eqref{eqn:cdf_nonrand_twoled_3}, which suggests both LEDs are out of sight. In this case channel gain is zero, which shows as a Dirac delta function at $x=0$ in pdf.

It is possible to simplify \eqref{eqn:cdf_nonrand_twoled_4} further if some particular scenarios of interest are assumed. These scenarios involving wide receiver FOV and non-negative incidence angles $\theta_i$'s are discussed with the simplified cdf expressions in the following remarks. 

\begin{remark} \label{rem:nonneg_theta}
When the receiver is pointing relatively upward and does not change this orientation dramatically, the incidence angles $\theta_1$ and $\theta_2$ take non-negative values only, which implies $\theta \,{>}\, 0$ and $\theta \,{<}\, \Phi$. In this particular case, the cdf in \eqref{eqn:cdf_nonrand_twoled_4} can be expressed by using the following simplified probabilities
\begin{align}
P_1(x) &= \bigtriangledown_\theta(z_1(x), \FOV, \max(0, \Phi {-} \FOV), \Phi {-} z_2(x)) \,, \nonumber\\
P_2(x) &= \bigtriangleup_\theta(z_1(x), \min(\FOV, \Phi \,{-}\, \FOV )) \,, \nonumber\\
P_3(x) &= \bigtriangledown_\theta(\Phi \,{-}\, \FOV, \Phi \,{-}\, z_2(x), \FOV, \Phi) \,,\nonumber\\
P_4 &= \bigtriangleup_\theta(\FOV, \Phi \,{-}\, \FOV)\,, \nonumber
\end{align}
which can be obtained directly from \eqref{eqn:p1}-\eqref{eqn:p4} by assuming the condition $0 \,{<}\, \theta \,{<}\, \Phi$ and following the respective derivation steps in Appendix~\ref{app:nonrand_twoled}. The pdf of this case can be obtained using \eqref{eqn:pdf_nonrand_twoled} and \eqref{eqn:cdf_theta_del}.
\end{remark}

\begin{remark} \label{rem:wide_fov}
When the receiver FOV is sufficiently large, the LEDs are always in the FOV, and the cdf in \eqref{eqn:cdf_nonrand_twoled_4} now becomes
\begin{align}
\!F_{h_{\rm eff}^2}(x) \!&= \Pr\{ z_1(x) < |\theta_1|, z_2(x) < |\theta_2| \} \,, \nonumber \\
&= \bigtriangleup_\theta(z_1(x),\Phi {-} z_2(x)) {+} F_\theta(\min({-}z_1(x), \Phi {-} z_2(x)))\,, \nonumber\\
& \quad {+}\, 1 \,{-}\, F_\theta(\max(z_1(x), \Phi \,{+}\, z_2(x))), \nonumber  
\end{align}
which is equivalent to \eqref{eqn:p1} with a large $\FOV$.
\end{remark}

\begin{remark} \label{rem:both}
When the incidence angles $\theta_1$ and $\theta_2$ take non-negative values and the receiver FOV is sufficiently wide, we have the following simplified cdf expression 
\begin{align}
F_{h_{\rm eff}^2}(x) &= \Pr\{ z_1(x) < \theta_1, z_2(x) < \theta_2 \} \,, \nonumber\\
&= \Pr\{ z_1(x) < \theta, z_2(x) < \Phi - \theta \} \nonumber \,, \\
&= \bigtriangleup_\theta(z_1(x),\Phi - z_2(x)) \,,  
\end{align}
which is equivalent to \eqref{eqn:p1} with $0 \,{<}\, \theta \,{<}\, \Phi$ and large $\FOV$. 
\end{remark}
The desired pdf expressions for the simplified cdf's in Remark~\ref{rem:wide_fov}-\ref{rem:both} can be computed using \eqref{eqn:cdf_theta_del} considering the detailed explanations in Appendix~\ref{app:nonrand_twoled}. Note that for these two remarks, wide FOV assumption is made. Therefore their cdfs are expected to be continuous, and their pdfs does not have a delta function.

\subsection{Two LEDs with Random User Location}
In the two LEDs case, the parts of the channel $h_{d_i}$ in \eqref{eqn:h_d_i} and $h_{\theta_i}$ in \eqref{eqn:h_theta_i} are now correlated through the geometric relation \eqref{eqn:theta_sum}. Therefore, when the horizontal distance $d$ is assumed to be random, it is not possible to derive the statistics using the identity employed in \eqref{eqn:cdf_joint} and \eqref{eqn:pdf_joint} for the product of the independent random variables. We therefore resort to a more conventional way of taking average over the distribution of the random distance $d$. As a result, once we obtain the distribution of the square-effective channel gain parametrically for a given $d$, the desired cdf and pdf can be calculated as
\begin{align}
F_{h_{\rm eff}^2}(x) = P_4 + \int_{\mathcal{R}_y} F_{h_{\rm eff}^2}(x|y)f_d(y) {\rm d}y,
\end{align}
\begin{align}
f_{h_{\rm eff}^2}(x) = 
\begin{cases}
c_h\int_{\mathcal{R}_y} f_{h_{\rm eff}^2}(x|y)f_d(y) {\rm d}y &{\mbox for~} x > 0 \\
P_4\delta(x) &{\mbox for~} x = 0
\end{cases},
\end{align}
respectively, where $f_d(\cdot)$ is the pdf of the random location $d$, and $\mathcal{R}_y$ is the support set.
\section{Channel Statistics for Multiple LEDs}\label{sec:multipleLED}

\begin{figure}[t]
\centering
\includegraphics[width=3.4in]{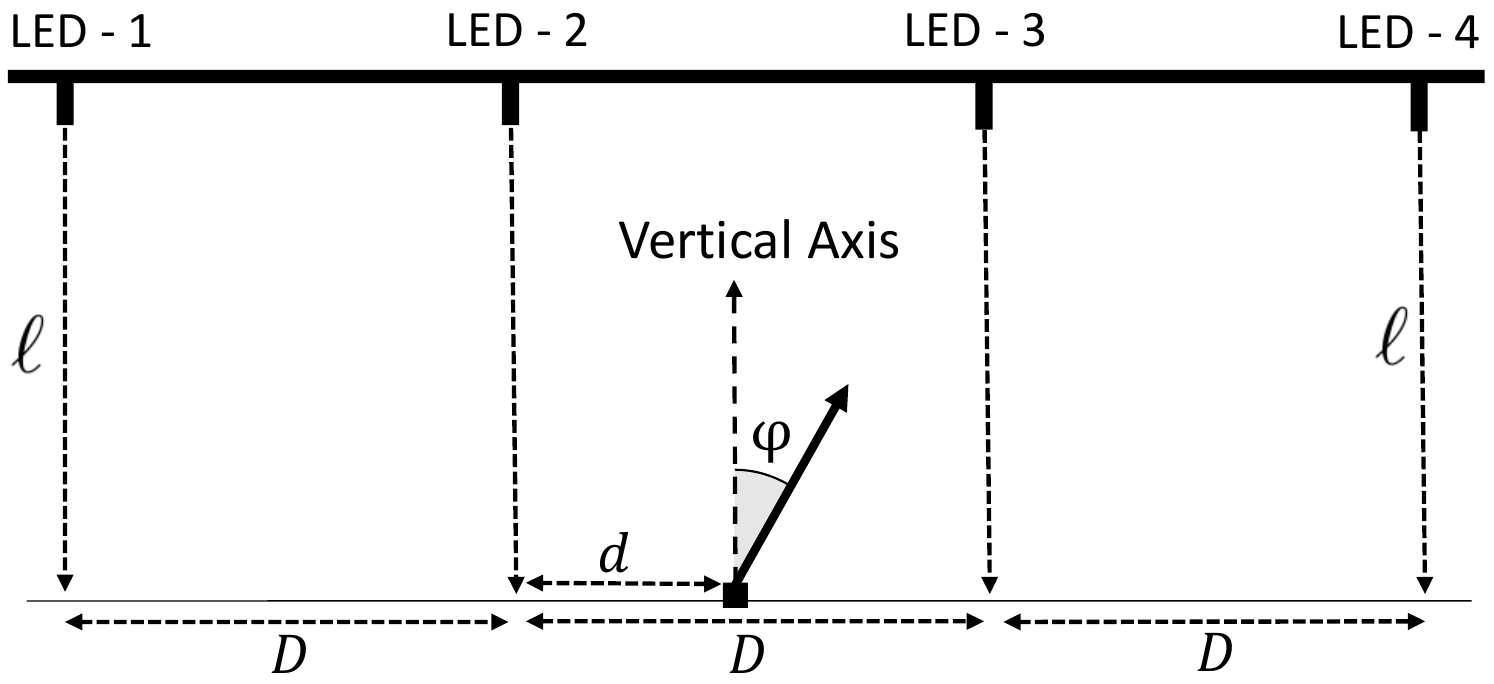}
\caption{A representative VLC downlink with $4$ LEDs deployed along a line with an equal spacing. The vertical orientation is characterized by the random angle $\varphi$.}
\label{fig:mulLED_setting}
\end{figure}

We have considered the single LED and two LEDs scenarios so far, and derived the square-channel distribution when the receiver orientation is random in the vertical direction. In this section, we consider a multiple LED scenario with more than two LEDS, and discuss the extension of the findings for the two LEDs case presented in Section~\ref{sec:binaryLED} to a multiple LED scenario. Our purpose here is not to provide a complete derivation for the channel statistics, but to consider two representative scenarios to gain some insight. 

\begin{figure}[t]
\centering
\includegraphics[width=0.49\textwidth]{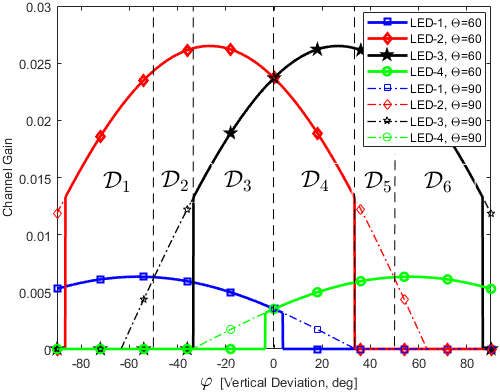}
\caption{Channel gains for $4$ LEDs of the configuration in Fig.~\ref{fig:mulLED_setting} with $D\,{=}\,3$~m, $\ell\,{=}\,3$~m, $d\,{=}\,1.5$~m, and $\FOV\,{=}\,\{60^\circ,90^\circ\}$.}
\label{fig:mulLED_midd}
\end{figure} 

We consider a representative multiple LED scenario with $4$ LEDs each of which are deployed along a line with equal spacing $D$, as shown in Fig.~\ref{fig:mulLED_setting}. The receiver is assumed to be located between two inner LEDs, which are labeled as LED-$2$ and LED-$3$, and is away from LED-$2$ by a distance $d$. Without any loss of generality, we assume that the receiver is facing upward with a deviation from the vertical axis by a random angle $\varphi$, and that $D\,{=}\,3$~m, and $\ell\,{=}\,3$~m. As before, we assume that the user is assigned to the LED with the strongest signal, which depends on the relative distances between the user and the LEDs, as well as the incidence angles. Note that, the effect of the incidence angle on the signal strength does not appear only due to the Lambertian pattern of the LED, but also because of the FOV evaluation, and both will be examined in the sequel.

We first assume that the user in Fig.~\ref{fig:mulLED_setting} is located in the middle of LED-$2$ and LED-$3$ with $d\,{=}\,1.5$~m, and that $\Phi_{1/2}\,{=}\,60^{\circ}$ and $\Ar\,{=}\,1~\textrm{cm}^2$. The associated signal qualities of the four LEDs at the receiver are depicted in Fig.~\ref{fig:mulLED_midd} with varying vertical deviation $\varphi$, and for two different FOV choices of $\FOV\,{=}\,\{60^\circ,90^\circ\}$. We partitioned Fig.~\ref{fig:mulLED_midd} into the regions $\mathcal{D}_i, ~i = 1,..,6,$ with respect to the LEDs having the first two strongest signal level among the others, and tabulate them in Table~\ref{tab:LEDpowers_midd} for each region. 

\begin{table}[!h]
\centering
\renewcommand{\arraystretch}{1.3}
\caption{LEDs with the First Two Strongest Signal for Fig.~\ref{fig:mulLED_midd}.}
\label{tab:LEDpowers_midd}
\begin{tabular}{|c|c|c|c|c|}
\hline
\multirow{2}{*}{Region} & \multicolumn{2}{|c|}{Wide FOV} & \multicolumn{2}{|c|}{Narrow FOV}\\
\cline{2-5}
	& Strongest & $2$nd-Strongest	& Strongest & $2$nd-Strongest \\
\hline\hline
$\mathcal{D}_1$ & LED-$2$	& LED-$1$	& LED-$2$	& LED-$1$ \\
$\mathcal{D}_2$ & LED-$2$	& LED-$3$	& LED-$2$	& LED-$1$ \\
$\mathcal{D}_3$ & LED-$2$	& LED-$3$	& LED-$2$	& LED-$3$ \\
$\mathcal{D}_4$ & LED-$3$	& LED-$2$	& LED-$3$	& LED-$2$ \\
$\mathcal{D}_5$ & LED-$3$	& LED-$2$	& LED-$3$	& LED-$4$ \\
$\mathcal{D}_6$ & LED-$3$	& LED-$4$	& LED-$3$	& LED-$4$ \\
\hline
\end{tabular}
\end{table}

We observe from Table~\ref{tab:LEDpowers_midd} that comparing the signal strength of the same two LEDs (LED-$2$ and LED-$3$) is sufficient to find the strongest of the $4$ LEDs in either $\mathcal{D}_2$-$\mathcal{D}_5$ for the wide FOV, or $\mathcal{D}_3$-$\mathcal{D}_4$ for the narrow FOV. This result implies that when the vertical deviation $\varphi$ falls into one of these group of regions, the findings of the two LEDs scenario in Section~\ref{sec:binaryLED} can be directly used to derive the channel statistics for this multiple LED case. Otherwise, the findings for the two LEDs scenario may need to be extended to the joint statistics of three or four LEDs. For example, $\varphi$ range corresponding to $\mathcal{D}_2$-$\mathcal{D}_5$ under the narrow FOV requires the joint cdf statistics of all $4$ LEDs  to find the strongest signal. In general, as the receiver FOV is relatively wide or the receiver orientation does not vary much, the two LEDs results apply for the multiple LED cases similar to the one in Fig.~\ref{fig:mulLED_setting}. 

\begin{figure}[t]
\centering
\includegraphics[width=0.49\textwidth]{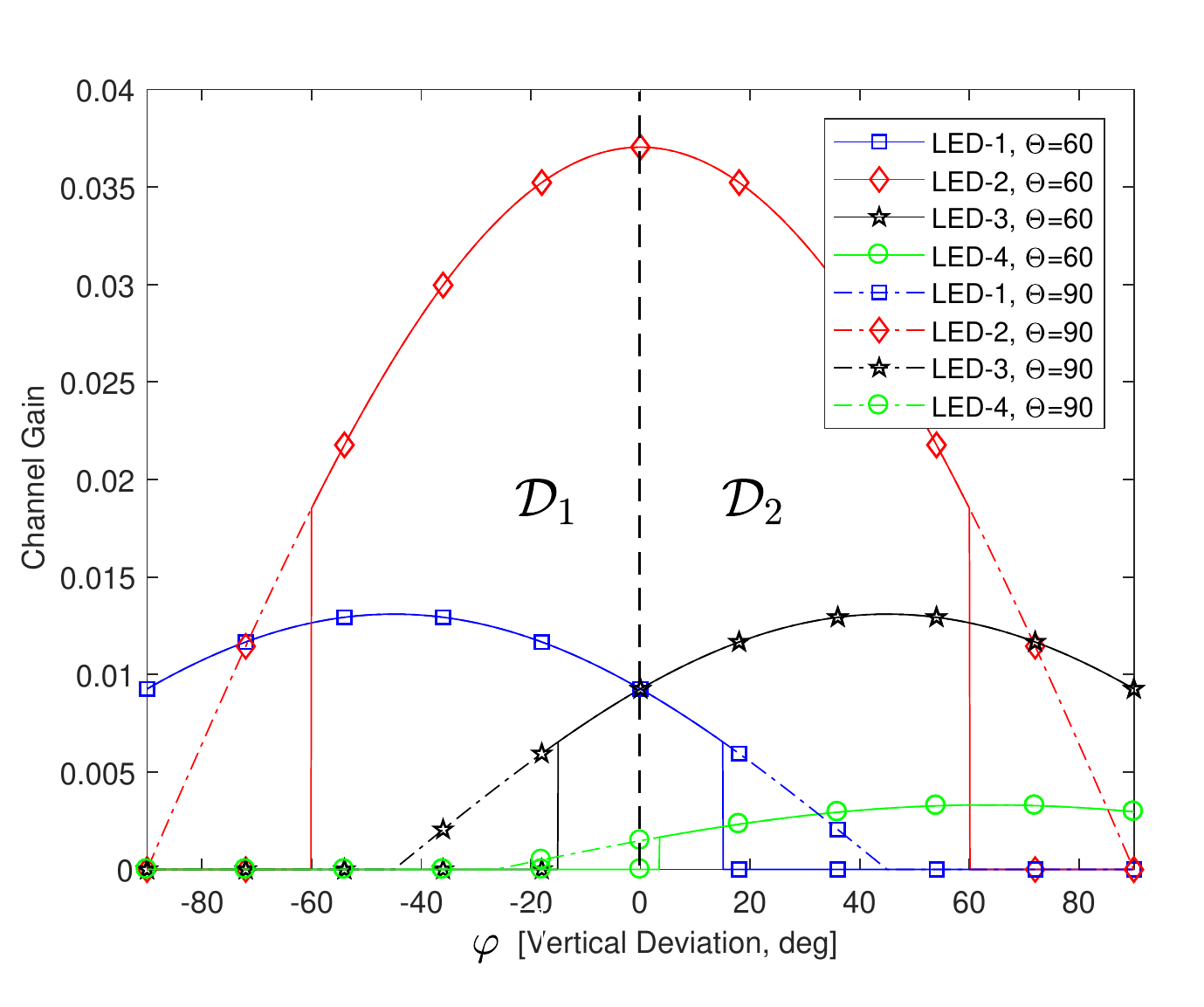}
\caption{Channel gains for $4$ LEDs of the configuration in Fig.~\ref{fig:mulLED_setting} with $D\,{=}\,3$~m, $\ell\,{=}\,3$~m, $d\,{=}\,0$~m, and $\FOV\,{=}\,\{60^\circ,90^\circ\}$.}
\label{fig:mulLED_left}
\end{figure} 

As a marginal example, we consider a second scenario where the user in Fig.~\ref{fig:mulLED_setting} is located directly under LED-$2$ with $d\,{=}\,0$. Based on the respective signal strength results depicted in Fig.~\ref{fig:mulLED_left}, the results of the two LEDs case directly applies if $\varphi$ takes either negative ($\varphi\,{\in}\,\mathcal{D}_1$) or positive values ($\varphi\,{\in}\,\mathcal{D}_2$), and otherwise requires an extension to the joint statistics of three LEDs. 

\section{Numerical Results}\label{sec:results}
In this section, we present the numerical results for the distribution of the square-channel with the single and two LEDs setting under various statistics for the random vertical orientation. In both LED settings, we provide numerical results for only the random choice of the horizontal distance $d$, and consider the relatively simple deterministic $d$ case as a subset. We assume that $D\,{=}\,4$~m, $\ell\,{=}\,3$~m, $\Phi_{1/2}\,{=}\,60^{\circ}$, $g\,=\,1$, and $\Ar\,{=}\,1$~cm$^2$, without any loss of generality. Like~\cite{Karagiannidis2017OnPerVLC}, we consider $E_s/N_0$ to characterize the transmit signal-to-noise ratio (SNR), and we study the impact of random receiver orientation and location on the BER and outage probability performance.  

\begin{figure}[!t]
\centering
\subfloat[{$\theta$ is uniformly distributed with $\mathcal{U}\,[20^{\circ},40^{\circ}]$, and $d$ is uniformly distributed with $\mathcal{U}[0,5]~\textrm{m}$}. The magnitude of the Dirac delta is $c_\delta\,{=}\,\{0.25,0\}$ for $\FOV\,{=}\,\{35^\circ,60^\circ\}$, respectively.]{\includegraphics[width=0.48\textwidth]{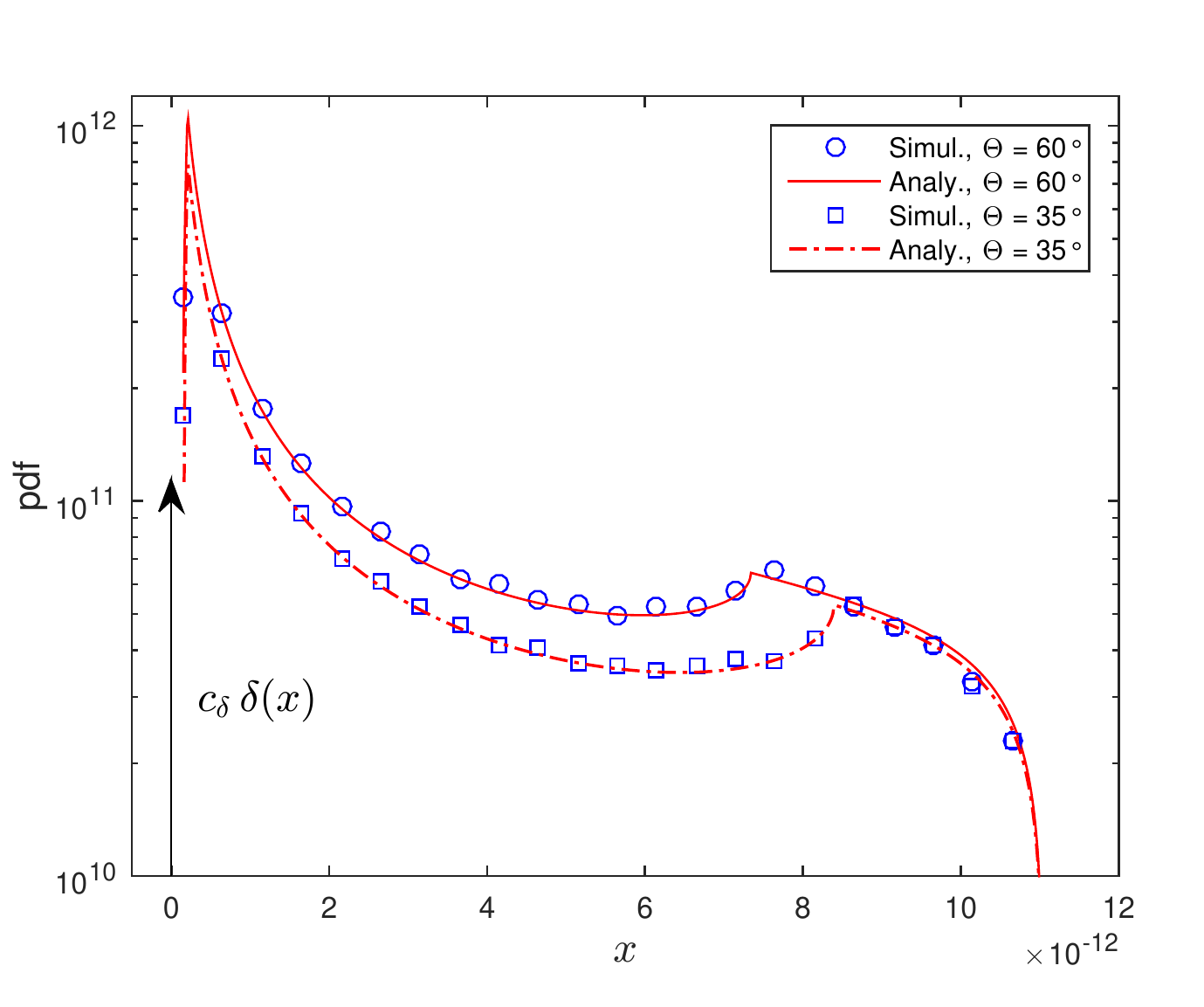}
\label{fig:pdf_eta_uniform_d_uniform}}\\
\subfloat[{$\theta$ is Gaussian distributed with $\mathcal{N}\,(30^{\circ},20^{\circ})$, and $d$ is Rayleigh distributed with $\mathcal{R}(1)~\textrm{m}$}. The magnitude of the Dirac delta is $c_\delta\,{=}\,\{0.1318,9.85{\times}10^{{-}12}\}$ for $\FOV\,{=}\,\{35^\circ,60^\circ\}$, respectively.]{\includegraphics[width=0.48\textwidth]{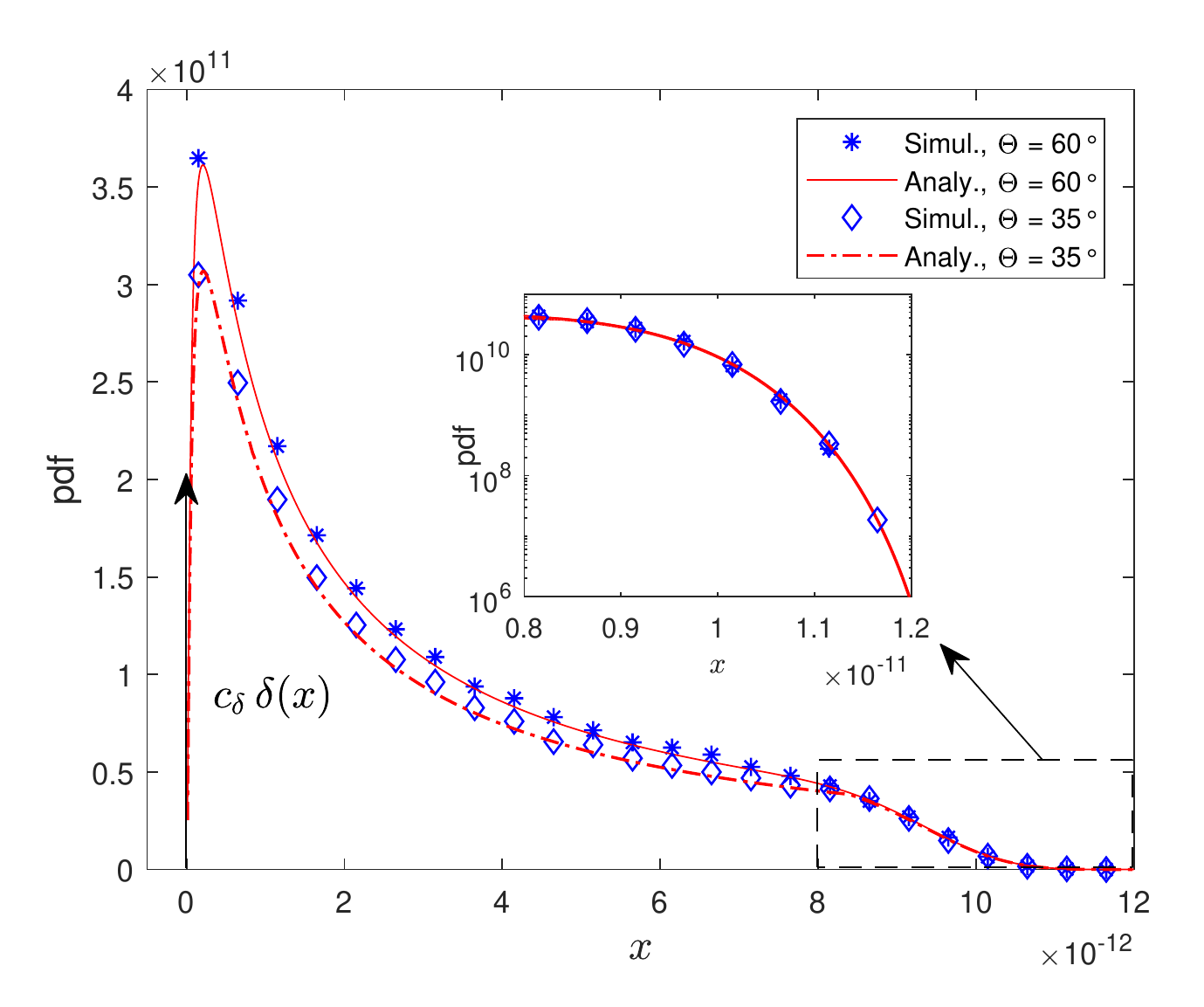}
\label{fig:pdf_eta_normal_d_rayleigh}}
\caption{The analytical and simulation data for the pdf of the square-channel in the single LED setting, where both the vertical orientation $\theta$ and the horizontal distance $d$ are random.}
\label{fig:pdf_singleLED}
\end{figure}

\begin{figure}[!t]
\centering
\includegraphics[width=0.46\textwidth]{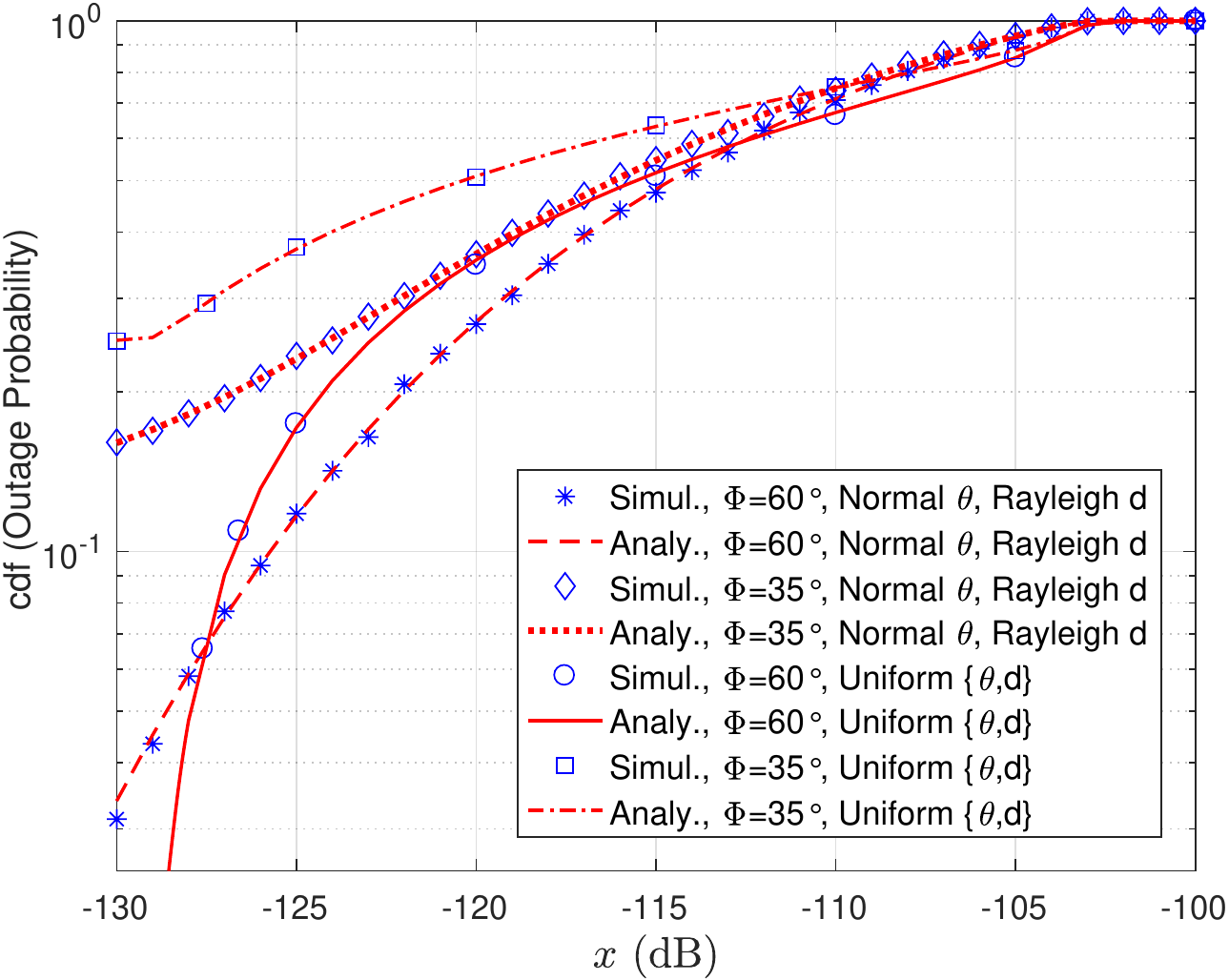}
\caption{The analytical and simulation data for the cdf of the square-channel in the single LED setting, where $\theta$ and $d$ follow the distributions in Fig.~\ref{fig:pdf_singleLED}.}
\label{fig:cdf_singleLED}
\end{figure}

\subsection{Single LED Case}
In Fig.~\ref{fig:pdf_singleLED}, the pdf of the square-channel $h^2$ is depicted for the single LED case, where both the incidence angle $\theta$ representing the vertical orientation of the user and the horizontal distance $d$ are random variables with various distributions. In particular, $\theta$ is assumed to follow uniform distribution with $\mathcal{U}\,[20^{\circ},40^{\circ}]$ or Gaussian distribution with $\mathcal{N}\,(30^{\circ},20^{\circ})$, both of which have the same mean value, whereas $\mathcal{U}[0,5]~\textrm{m}$ or $\mathcal{R}(2)~\textrm{m}$ are assumed for $d$, both of which have similar mean values. 

We observe a perfect match between the analytical and simulation data of Fig.~\ref{fig:pdf_singleLED} for all distributions of $\theta$ and $d$, which verifies the analytical derivation of the square-channel distribution. Note that, the FOV value of $\FOV\,{=}\,60^\circ$ covers the incidence angle $\theta$ range completely for $\mathcal{U}\,[20^{\circ},40^{\circ}]$, and with a very high probability for $\mathcal{N}\,(30^{\circ},20^{\circ})$. On the other hand, the relatively narrow FOV value of $\FOV\,{=}\,35^\circ$ does not cover the angle span of $\theta$ completely in either case. As a result, the Dirac delta $\delta(x)$, which arises from the FOV values smaller than the angle span of $\theta$, does not appear at all or appears with a very small magnitude $c_\delta$ for $\FOV\,{=}\,60^\circ$, whereas we have $c_\delta\,{=}\,\{0.25,0.1318\}$ for $\mathcal{U}\,[20^{\circ},40^{\circ}]$ and $\mathcal{N}\,(30^{\circ},20^{\circ})$, respectively, for $\FOV\,{=}\,35^\circ$. These results exemplify the mechanism how the narrowing FOV introduces a nonzero probability of the LOS communication link being out of the FOV, which in turn causes a Dirac delta at the pdf of the square-channel, as explained in Section~\ref{sec:singleLED}. We also observe that the pdf statistics of different FOV configurations are almost the same after relatively large input values along $x$-axis. This is because the high channel gains can be observed when the receiver orientation is well aligned ($\theta$ is small), and in this case LED being out of FOV is a small probability bot FOV configurations. The respective cdf statistics for Fig.~\ref{fig:pdf_singleLED} are provided in Fig.~\ref{fig:cdf_singleLED}, which also correspond to outage probability for a given channel gain requirement. Outage probability difference can be most explicitly observed for lower channel gain requirement values, where we can observe up to ten times higher outage probability with low FOV receivers. 

\begin{figure}[!t]
\centering
\includegraphics[width=0.47\textwidth]{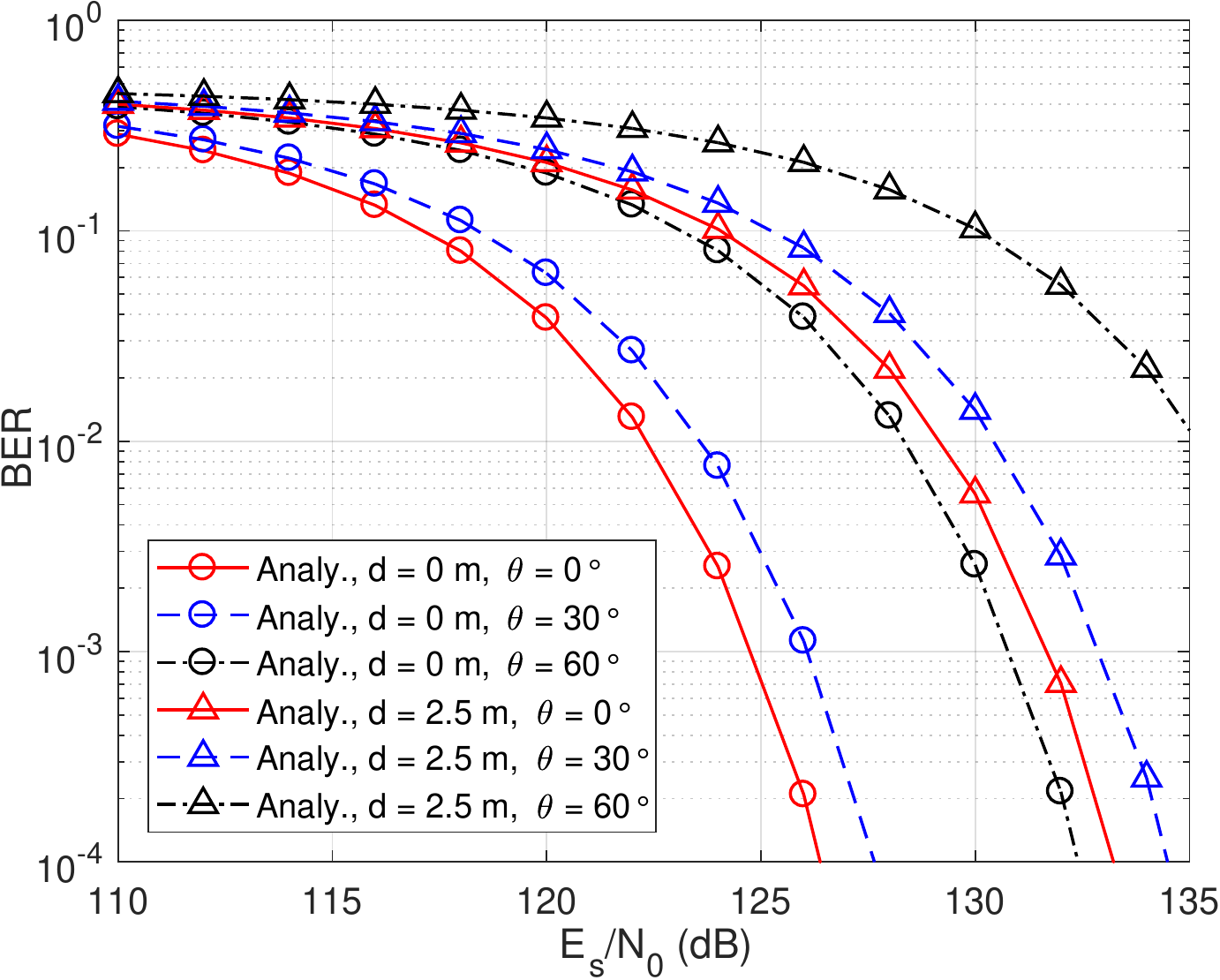}
\caption{BER results for the single LED setting with non-mobile non-fluctuation (fixed) scenarios with different $\theta$ and $d$ values for $\FOV \ge 60^\circ$.}
\label{fig:BERFixed}
\end{figure}

\begin{figure}[!t]
\centering
\includegraphics[width=0.48\textwidth]{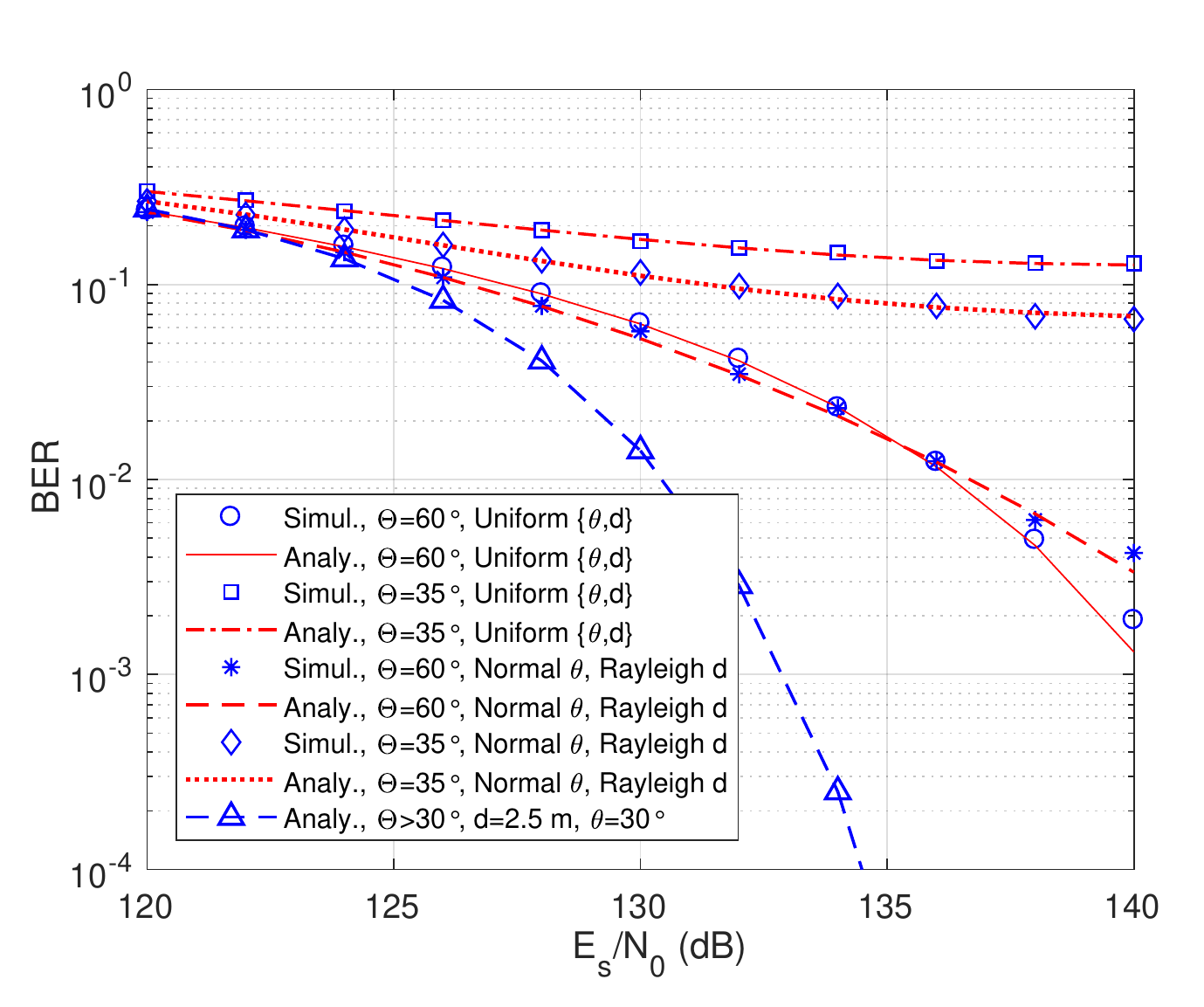}
\caption{BER results for the single LED setting with the distributions of $\theta$ and $d$ considered in Fig.~\ref{fig:pdf_singleLED}, as well as a non-mobile non-fluctuation (fixed) scenario for comparison.}
\label{fig:ber_singleLED}
\end{figure}

Before providing the BERs for given channel pdfs, we provide the BERs of some non-mobile and fixed user orientation scenarios in Fig.~\ref{fig:BERFixed} to give intuition about the effect of horizontal distance and orientation on the BER and provide benchmarks for later results. The FOV is assumed to be larger than $\theta$ in all scenarios. In order to achieve lower BERs, more than $120$~dB $E_s/N_0$ is required for all scenarios\footnote{For a representative study on typical transmit SNR range for indoor deployment of multi-LED VLC systems, the reader is referred to~\cite{7362097}.}. The scenario with $\theta = 0^\circ$, and $d = 0$~m is the case when the receiver is located immediately underneath and facing the LED. This scenario provides the highest possible channel gain, thus lowest possible BER. The orientation change of $30^\circ$ increases BER slightly, while a change of $60^\circ$ causes a much higher BER. On the other hand, horizontal distance of 2.5 m causes a significant BER increase too.

While Fig.~\ref{fig:BERFixed} illustrates BERs for fixed location and orientation scenarios, Fig.~\ref{fig:ber_singleLED} illustrates the BERs for random location and orientation scenarios with the channel pdfs shown in Fig.~\ref{fig:pdf_singleLED}. A representative scenario with non-mobile and fixed user orientation is also provided, where $\theta = 30^\circ$ and $d = 2.5$~m. Note that all scenarios in Fig.~\ref{fig:ber_singleLED} has the same mean $\theta$ and approximately same mean horizontal distance. The random fluctuations increase the BER significantly compared to the fixed setting with the same mean $\theta$ and $d$. Although having much higher BER than the fixed setting, the curve of the relatively wide FOV value of $\FOV\,{=}\,60^\circ$ show monotonically decaying characteristics. On the other hand, the narrow FOV of $\FOV\,{=}\,35^\circ$ cause the BER curves to saturate, where the saturation value is related to the probability of the LOS link being out of the FOV, which is equal to $c_\delta$ shown in Fig.~\ref{fig:pdf_singleLED}. Since the error probability of OOK is 0.5 for zero channel gain, the BERs saturate to $c_\delta/2$ for each case. This result shows that a reliable communication is not possible when there is a high possibility of LED being out of FOV. This problem can be solved by increasing the FOV of the receiver, or deploying more LEDs to make sure receiver is in contact to at least one LED in FOV.

\begin{figure}[!t]
\centering
\subfloat[{$\theta$ is uniformly distributed with $\mathcal{U}\,[20^{\circ},40^{\circ}]$, and $d$ is uniformly distributed with $\mathcal{U}[0,D]~\textrm{m}$}.]{\includegraphics[width=0.45\textwidth]{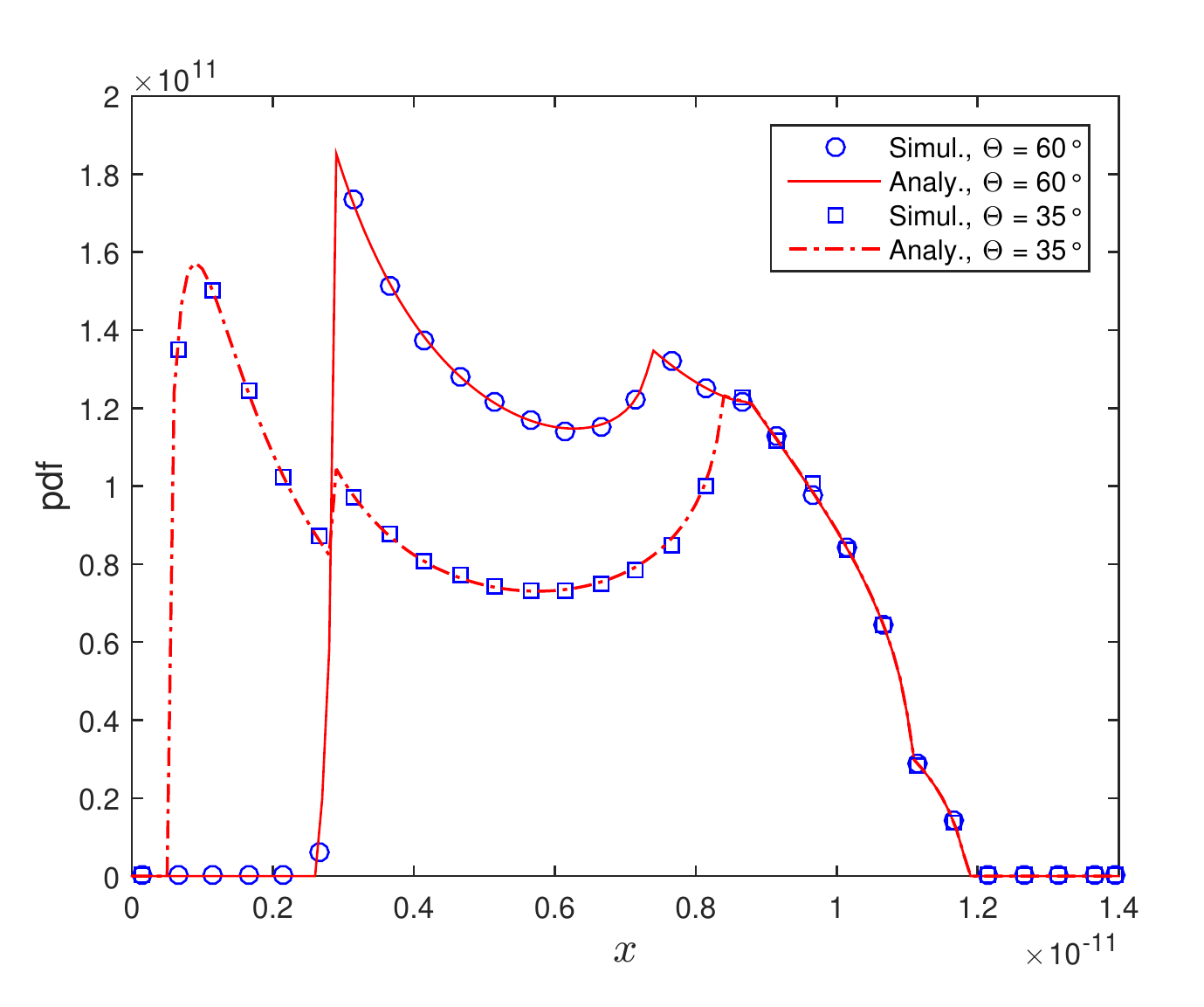}
\label{fig:pdf_2LEDsUniform}}\\
\subfloat[{$\theta$ is Gaussian distributed with $\mathcal{N}\,(30^{\circ},20^{\circ})$, and $d$ is uniformly distributed with $\mathcal{U}[0,D]~\textrm{m}$}.]{\includegraphics[width=0.45\textwidth]{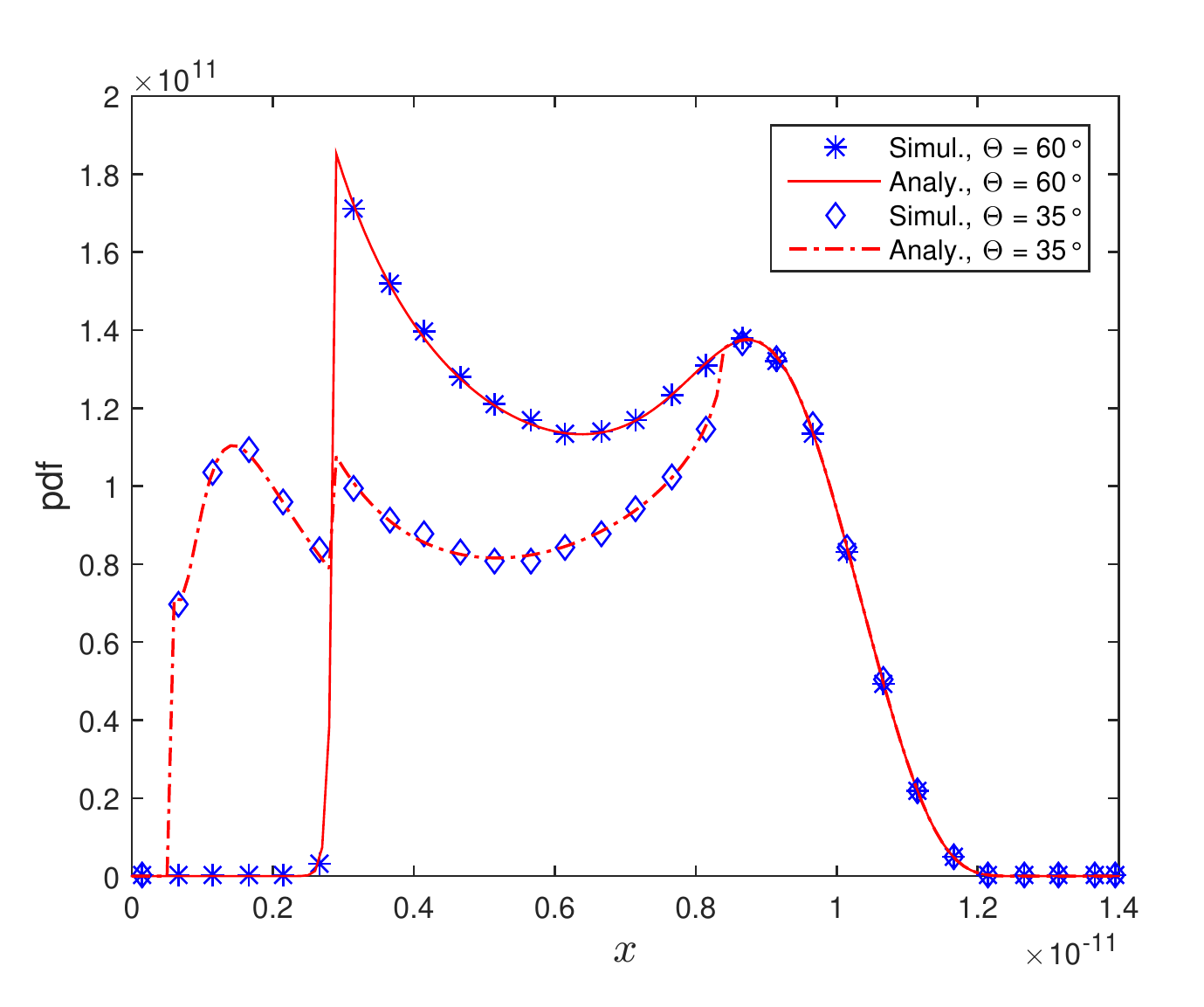}
\label{fig:pdf_2LEDsNormal}}
\caption{The analytical and simulation data for the pdf of the square-channel in the two LEDs setting, where both the vertical orientation $\theta$ and the horizontal distance $d$ are random.}
\label{fig:pdf_binaryLED}
\end{figure}

\begin{figure}[!t]
\centering
\includegraphics[width=0.47\textwidth]{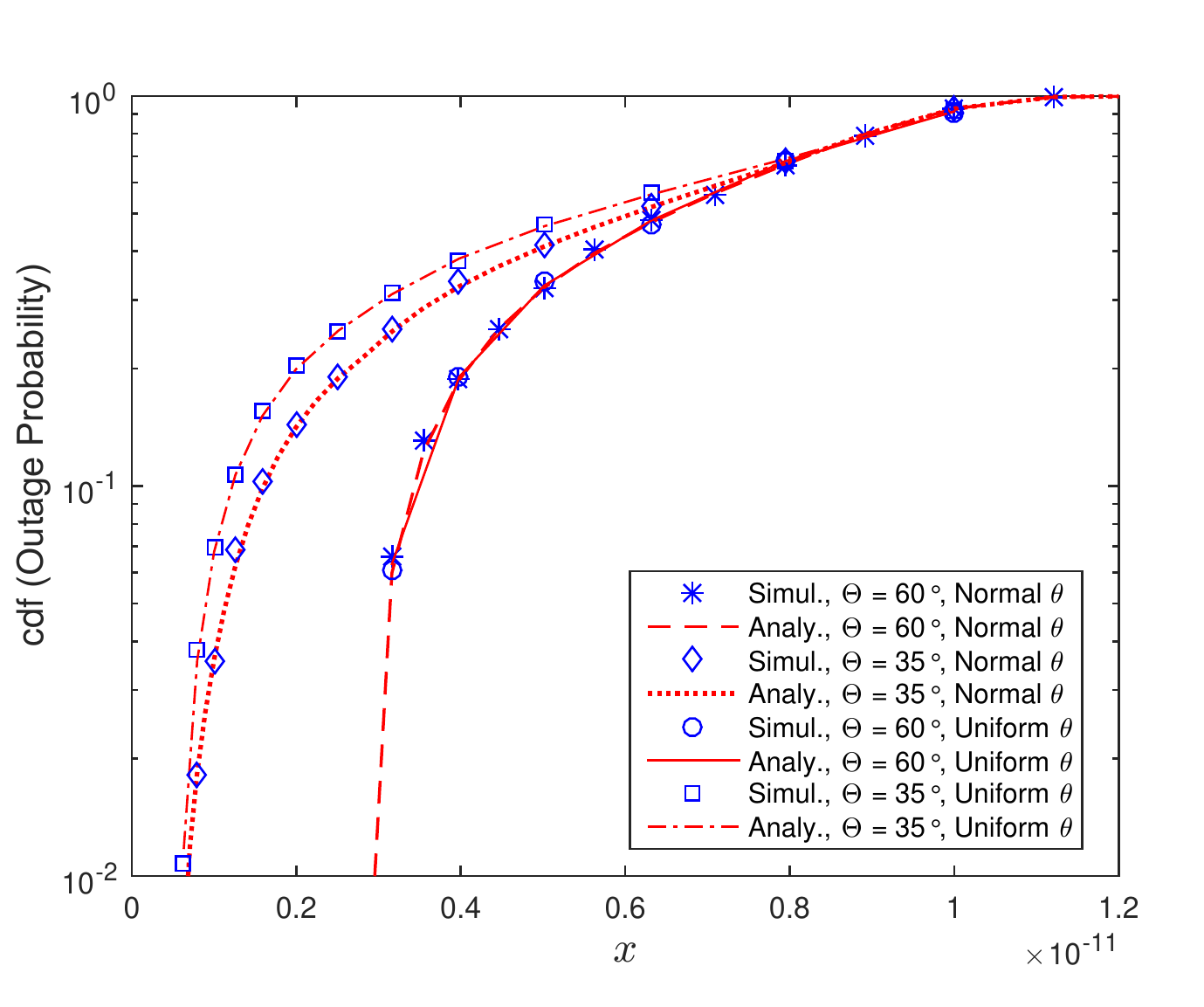}
\caption{The analytical and simulation data for the cdf of the square-channel in the two LEDs setting, where $\theta$ and $d$ follow the distributions in Fig.~\ref{fig:pdf_binaryLED}.}
\label{fig:cdf_binaryLED}
\end{figure}

\begin{figure}[!t]
\centering
\includegraphics[width=0.5\textwidth]{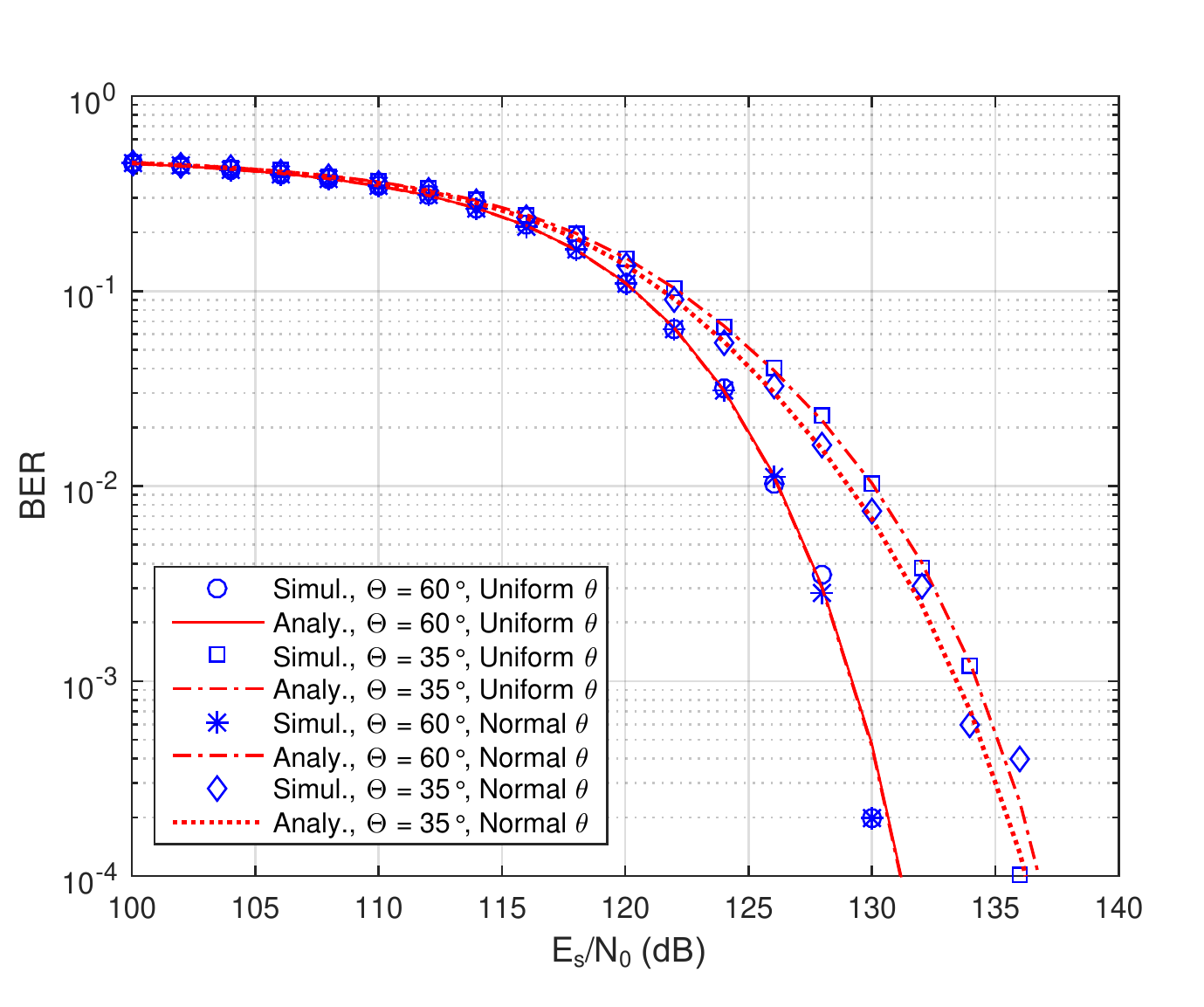}
\caption{BER results for the two LEDs setting, where $\theta$ follows the distributions in Fig.~\ref{fig:pdf_binaryLED}, and $d\,{\sim}\,\mathcal{U}[0,D]~\textrm{m}$.}
\label{fig:ber_binaryLED}
\end{figure}

\subsection{Two LEDs Case}
The pdf of the square-effective channel gain for the two LEDs setting is presented in Fig.~\ref{fig:pdf_binaryLED} under various distributions for $\theta$ and $d$. In particular, we assume $\mathcal{U}\,[20^{\circ},40^{\circ}]$ and $\mathcal{N}\,(30^{\circ},20^{\circ})$ for $\theta$, as before, and $\mathcal{U}\,[0,D]$~m for $d$. We observe that the analytical results follow the simulation data successfully. In this two LEDs setting, since the LEDs are never out of the receiver FOV simultaneously for the given FOV values of $\FOV\,{=}\,\{35^\circ,60^\circ\}$, there is no possibility in which the effective channel gain becomes zero, and we therefore do not have any Dirac delta appearing in Fig.~\ref{fig:pdf_binaryLED}. We observe that as the FOV becomes smaller, the pdf curve broadens towards the origin and covers much smaller $x$ values. Smaller receiver FOV angles increase the possibility of the LOS links being out of the FOV. When stronger link is blocked due to FOV limit, the receiver connects to the LED with weaker link, causing the effective channel gain taking much smaller values. Note that even more smaller FOV angles will eventually cause the broadened pdf to include the origin, which will accordingly appear as the Dirac delta at the origin, as in the single LED scenario. The respective cdf statistics (outage probabilities) are also provided in Fig.~\ref{fig:cdf_binaryLED}. 

We finally depict the BER results of the two LEDs setting in Fig.~\ref{fig:ber_binaryLED}. Since the user does not lose contact with both the LEDs simultaneously, the effective channel gain is always nonzero, and the BER statistics do not saturate at all. Although the narrower FOV of $\FOV\,{=}\,35^\circ$ deteriorate the BER performance, it is better than that for the single LED case depicted in Fig.~\ref{fig:ber_singleLED} with the same FOV value. This result highlights the multiple LED deployment~\cite{Eroglu_JSAC} as an effective direction to cope with adverse effects of the random orientation.

\section{Conclusion}\label{sec:conclusion}
We investigated the statistics of a VLC downlink channel when the orientation of the user is varying randomly around the vertical axis. The mobility is also considered through the random deployment of the user which results in a random distance to the source LED. We observe that the random fluctuations in the vertical user orientation can adversely affect the achievable user data rate. We proposed an analytical framework which successfully characterizes the channel statistics when both the vertical orientation and the user location are randomly varying. This analytical framework serves as a practical basis, as well, to develop strategies in dealing with the destructive effects of the random vertical orientation with random user location. Our results show that for a receiver horizontally located 2.5 meter away from the transmitter with an orientation angle of $30^\circ$, random deviations in the receiver orientation/location results in more than 7~dB of SNR loss at a BER of $10^{-3}$ dB for a wide FOV. For a narrow FOV, the effects are even more catastrophic, where the BER quickly converges to an error floor as the transmit power increases. As a future research direction, the analysis for the random orientation around the vertical $y$ axis can be extended to cover the other two Euler angles defining the rotation around the $x$ and $z$ axes, as well.

\appendices
\section{Proof of Lemma~\ref{lem:delta_two_int}} \label{app:delta_two_int}
\begin{figure}[t]
\centering
\includegraphics[width=0.4\textwidth]{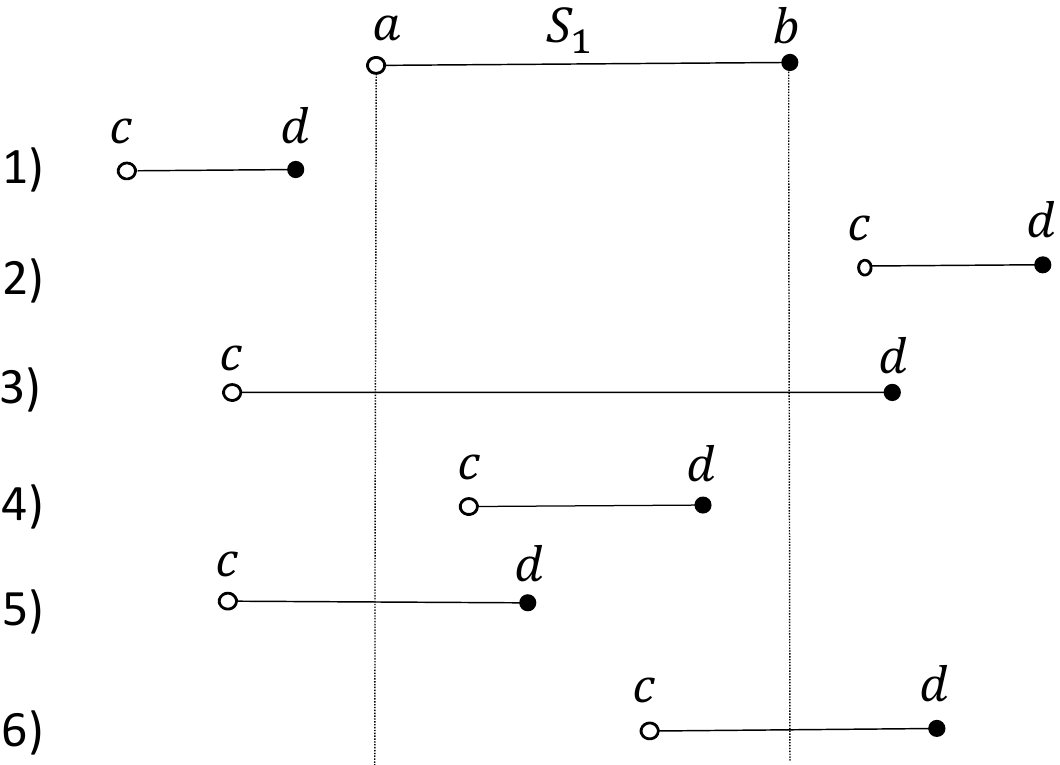}
\caption{Different intersection possibilities of the non-empty support sets $\mathcal{S}_1\,{=}\,(a, b\,]$ and $\mathcal{S}_2\,{=}\,(c, d\,]$.}
\label{fig:intersection}
\end{figure}

In \eqref{eqn:delta_two_int_1}, there are two support intervals of $\theta$, which are $\mathcal{S}_1\,{=}\,(a, b\,]$ and $\mathcal{S}_2\,{=}\,(c, d\,]$, and the function $\bigtriangledown_\theta(a, b, c, d)$ is equivalent to $\Pr\{ \theta \in \mathcal{S}_1 \cap \mathcal{S}2 \}$. In order to have a nonzero value for this joint probability, the conditions $a\,{\leq}\,b$ and $c\,{\leq}\,d$ have to be satisfied simultaneously, which guarantees that the support intervals are non-empty sets. In such a case, there are six different possibilities for the intersection of $\mathcal{S}_1$ and $\mathcal{S}_2$, as shown in the illustration of Fig.~\ref{fig:intersection}.

Realizing that for the first two cases in Fig.~\ref{fig:intersection}, represented with the labels $(1)$ and $(2)$, the intersection of the support sets is empty, i.e., $\mathcal{S}_1\,{\cap}\,\mathcal{S}_2\,{=}\,\varnothing$. In order to circumvent these two cases, we introduce two additional conditions $d\,{>}\,a$ and $c\,{<}\,b$ to be satisfied. As a result, all the conditions for non-empty intersection can be written jointly as
\begin{align}\label{eqn:nonempty_intersect}
\left\lbrace a,c \,|\, a < \min(b,d), \, c < \min(b,d)\right\rbrace \,.
\end{align}

For the cases $(3){-}(6)$ in Fig.~\ref{fig:intersection}, the non-empty intersections are given as follows
\begin{align}\label{eqn:intersection}
\mathcal{S}_1 \cap \mathcal{S}_2 &= \begin{cases}
(a,b\,] &\mbox{for } c \,{\leq}\, a, d \,{>}\, b 	\quad (\text{Case }3)  \\
(c,d\,] &\mbox{for } c \,{>}\, a , d \,{\leq}\, b 	\quad (\text{Case }4)  \\
(a,d\,] &\mbox{for } c \,{\leq}\, a, d \,{\leq}\, b \quad (\text{Case }5)  \\
(c,b\,] &\mbox{for } c \,{>}\, a , d \,{>}\, b		\quad (\text{Case }6)  \\
\end{cases},
\end{align}
which directly comes from the geometrical comparison of the support sets $\mathcal{S}_1\,{=}\,(a, b\,]$ and $\mathcal{S}_2\,{=}\,(c, d\,]$. Because $\Pr\{\theta \in (x,y\,]\}\,{=}\,F_\theta(y)\,{-}\,F_\theta(x)$, we readily obtain \eqref{eqn:delta_two_int_2} using \eqref{eqn:intersection} and the the cdf function $F_\theta(\cdot)$. Note that, the cases in \eqref{eqn:intersection} satisfy the non-empty intersection condition in \eqref{eqn:nonempty_intersect}.

\section{Proof of Theorem~\ref{the:cdf_nonrand_twoled}} \label{app:nonrand_twoled}
In order to prove the cdf expression in Theorem~\ref{the:cdf_nonrand_twoled}, we first represent the events defined in \eqref{eqn:event} by using the transformation $z_i(x)\,{=}\,\frac{1}{2}\cos^{{-}1}(2\frac{x}{c_i}{-}1)$ for $i\,{=}\,1,2$ as follows 
\begin{align}\label{eqn:event_z}
\mathcal{E}_{z_i} : \left\lbrace z_i(x) \in \Omega_{z_i} \,|\, z_i(x) < |\theta_i| \right\rbrace \,,
\end{align}
where $\Omega_{z_i}$ is the sample space of $z_i(x)$, and we employ the fact that $\cos(\cdot)$ is an even function. In the following, we separately derive the expressions for the probabilities corresponding to each of the four cases in \eqref{eqn:cdf_nonrand_twoled_3}. Note that, because all these four cases represent disjoint events, the desired probability \eqref{eqn:cdf_nonrand_twoled_4} is readily obtained by summing up their respective probability expressions via the law of total probability~\cite{Papoulis2002Prob}.  

As a remark, since all of the four cases in \eqref{eqn:cdf_nonrand_twoled_3} have inequalities of the form $|\theta_i| \,{\lessgtr}\, \FOV$, which have the roots $\theta \,{=}\, \{ 0, \FOV\}$, we analyze each probability of interest by considering the three disjoint support regions for the random angle $\theta$ given as
\begin{align}
\mathcal{S}^i_\theta = \begin{cases}
({-}\infty,0] &\mbox{for } i\,{=}\,1 \\
(0,\FOV] &\mbox{for } i\,{=}\,2 \\
(\FOV,\infty) &\mbox{for } i\,{=}\,3 \\
\end{cases}\,.
\end{align}
\begin{case}
Consider $P_1(x)\,{=}\,\Pr\left\lbrace \mathcal{E}_{z_1},\mathcal{E}_{z_2}, |\theta_1| \,{\leq}\, \FOV, |\theta_1| \,{\leq}\, \FOV \right\rbrace$. 
\begin{align}
P_1(x) &= \Pr\left\lbrace z_1(x) \,{<}\, |\theta| \,{<}\, \FOV,\; z_2(x) \,{<}\, |\Phi \,{-}\, \theta| \,{<}\, \FOV \right\rbrace \label{eqn:app:p1_1}\\
&= \displaystyle\sum_{i=1}^{3} P_1\left(x, \theta \,{\in}\, \mathcal{S}^i_\theta \right)\label{eqn:app:p1_2}
\end{align}
where the individual probabilities $P_1\left(x, \theta \,{\in}\, \mathcal{S}^i_\theta \right)$'s are computed with the help of the function $\bigtriangledown_\theta(\cdot)$ in \eqref{eqn:delta_two_int_2} as follows
\begin{align}
& P_1\left(x,\theta \,{\in}\, \mathcal{S}^1_\theta \right) \nonumber\\
& \quad = \Pr\{ {-}\FOV \,{\leq}\, \theta \,{<} {-}z_1(x), \Phi {-} \FOV \,{<} \theta \,{\leq}\, \Phi \,{-}\, z_2(x), \theta \,{<}\, 0 \} \,, \nonumber \\
& \quad =\Pr\{ {-}\FOV \,{\leq}\, \theta \,{<} {-}z_1(x), \Phi {-} \FOV \,{<} \theta \,{\leq}\, \Phi\,{-}z_2(x) \} \,, \nonumber\\
& \quad = \bigtriangledown_\theta({-}\FOV, {-}z_1(x), \Phi \,{-}\, \FOV, \Phi\,{-}\,z_2(x)). \label{eqn:app:p11}
\end{align}
\begin{align}
& P_1\left(x,\theta \,{\in}\, \mathcal{S}^2_\theta \right) \nonumber\\
& \quad =\Pr\{z_1(x) \,{<}\, \theta \,{\leq}\, \FOV, \Phi \,{-}\, \FOV \,{<}\, \theta \,{\leq}\, \Phi \,{-}\, z_2(x), 0 \,{<}\, \theta \,{<}\, \Phi \} \,,\nonumber \\
& \quad =\Pr\{ z_1(x) \,{<}\, \theta \,{\leq}\, \FOV, \max(0, \Phi \,{-}\, \FOV) \,{<}\, \theta \,{\leq}\, \Phi \,{-}\, z_2(x) \} \,,\nonumber \\
& \quad =\bigtriangledown_\theta(z_1(x), \FOV, \max(0, \Phi {-} \FOV), \Phi {-} z_2(x)).\label{eqn:app:p12}
\end{align}
\begin{align}
& P_1\left(x,\theta \,{\in}\, \mathcal{S}^3_\theta \right) \nonumber\\
& \quad =\Pr\{ z_1(x) \,{<}\, \theta \,{\leq}\, \FOV, \Phi \,{+}\, z_2(x) \,{<}\, \theta \,{\leq}\, \Phi \,{+}\, \FOV , \theta \,{>}\, \Phi \} \,,\nonumber \\
&\quad =\Pr\{ z_1(x) \,{<}\, \theta \,{\leq}\, \FOV, \Phi \,{+}\, z_2(x) \,{<}\, \theta \,{\leq}\, \Phi \,{+}\, \FOV \} \,,\nonumber \\
& \quad = \bigtriangledown_\theta(z_1(x), \FOV, \Phi \,{+}\, z_2(x), \Phi \,{+}\, \FOV).\label{eqn:app:p13}
\end{align}
As a result, \eqref{eqn:app:p1_2} with \eqref{eqn:app:p11}-\eqref{eqn:app:p13} readily yields \eqref{eqn:p1}. \hfill\IEEEQEDhere
\end{case}  

\begin{case}
Consider $P_2(x)\,{=}\,\Pr\left\lbrace \mathcal{E}_{z_1}, |\theta_1| \,{\leq}\, \FOV, |\theta_2| \,{>}\, \FOV \right\rbrace$. 
\begin{align}
P_2(x) =& \Pr\{ z_1(x) \,{<}\, |\theta| \,{\leq}\, \FOV, |\Phi \,{-}\, \theta| \,{>}\, \FOV \} \label{eqn:app:p2_1}\\
&= \displaystyle\sum_{i=1}^{3} P_2\left(x,\theta \,{\in}\, \mathcal{S}^i_\theta \right)\label{eqn:app:p2_2}
\end{align}
where $P_2\left(x,\theta \,{\in}\, \mathcal{S}^i_\theta \right)$'s are computed as follows
\begin{align}
P_2\left(x,\theta \,{\in}\, \mathcal{S}^1_\theta \right) &= \Pr\{ {-}\FOV < \theta \,{<} {-}z_1(x), \theta \,{<}\, \min(0,\Phi {-} \FOV) \}\,, \nonumber \\
&= \Pr\{ {-}\FOV \,{<}\, \theta \,{\leq}\, \min({-}z_1(x), \Phi {-} \FOV) \}\,, \nonumber \\ 
&= \bigtriangleup_\theta({-}\FOV, \min({-}z_1(x), \Phi {-} \FOV)) \,.  \label{eqn:app:p21} 
\end{align}
\begin{align}
P_2\left(x,\theta \,{\in}\, \mathcal{S}^2_\theta \right) &= \Pr\{ z_1(x) \,{<}\, \theta \,{<}\, \FOV, 0 \,{<}\, \theta \,{\leq}\, \Phi \,{-}\, \FOV \}\,, \nonumber\\
&= \Pr\{ z_1(x) \,{<}\, \theta \,{<}\, \min(\FOV, \Phi \,{-}\, \FOV ) \}\,, \nonumber\\
&= \bigtriangleup_\theta(z_1(x), \min(\FOV, \Phi \,{-}\, \FOV )) \,. \label{eqn:app:p22} 
\end{align}
\begin{align}
P_2\left(x,\theta \,{\in}\, \mathcal{S}^3_\theta \right) &= \Pr\{ z_1(x) \,{<}\, \theta \,{<}\, \FOV, \theta \,{>}\, \Phi \,{+}\, \FOV \} \nonumber\\ &= 0 \,. \label{eqn:app:p23} 
\end{align}
Similarly, \eqref{eqn:app:p2_2} with \eqref{eqn:app:p21}-\eqref{eqn:app:p23} yields \eqref{eqn:p2}. \hfill\IEEEQEDhere
\end{case} 

\begin{case}  
Consider $P_3(x)\,{=}\,\Pr\left\lbrace \mathcal{E}_{z_2}, |\theta_1| \,{>}\, \FOV, |\theta_2| \,{\leq}\, \FOV \right\rbrace$.
\begin{align}
P_3(x) =&  \Pr\{ z_2(x) \,{<}\, |\Phi \,{-}\, \theta| \,{\leq}\, \FOV, |\theta| \,{>}\, \FOV \} \label{eqn:app:p3_1}\\
&= \displaystyle\sum_{i=1}^{3} P_3\left(x,\, \mathcal{S}^i_\theta \right)\label{eqn:app:p3_2}
\end{align}
where $P_3\left(x,\theta \,{\in}\, \mathcal{S}^i_\theta \right)$'s are given as follows
\begin{align}
P_3\left(x,\theta \,{\in}\, \mathcal{S}^1_\theta \right) &= \Pr\{ \Phi \,{-}\, \FOV \,{<}\, \theta \,{<}\, \Phi \,{-}\, z_2(x), \theta \,{<}\,{-}\FOV \} \nonumber\\ &= 0 \,. \label{eqn:app:p31} 
\end{align}
\begin{align}
P_3\left(x,\theta \,{\in}\, \mathcal{S}^2_\theta \right) &= \Pr\{ \Phi \,{-}\, \FOV \,{<}\, \theta \,{<}\, \Phi \,{-}\, z_2(x), \FOV \,{<}\,  \theta \,{<}\, \Phi \} \,, \nonumber \\
&= \bigtriangledown_\theta(\Phi \,{-}\, \FOV, \Phi \,{-}\, z_2(x), \FOV, \Phi) \,. \label{eqn:app:p32} 
\end{align}
\begin{align}
P_3\left(x,\theta \,{\in}\, \mathcal{S}^3_\theta \right) &= \Pr\{ \Phi \,{+}\, z_2(x) \,{<}\, \theta \,{<}\, \Phi \,{+}\, \FOV, \theta \,{>}\, \FOV, \theta \,{>}\, \Phi \} \,,\nonumber \\
&= \Pr\{ \max(\Phi \,{+}\, z_2(x), \FOV) \,{<}\, \theta \,{\leq}\, \Phi \,{+}\, \FOV \} \,,\nonumber \\
&= \bigtriangleup_\theta(\max(\Phi \,{+}\, z_2(x), \FOV), \Phi \,{+}\, \FOV) \,. \label{eqn:app:p33} 
\end{align}
As before, \eqref{eqn:app:p3_2} with \eqref{eqn:app:p31}-\eqref{eqn:app:p33} yields \eqref{eqn:p3}. \hfill\IEEEQEDhere
\end{case}  

\begin{case}  
$P_4\,{=}\, \Pr\left\lbrace |\theta_1| \,{>}\, \FOV, |\theta_2| \,{>}\, \FOV \right\rbrace$.
\begin{align}
P_4 &= \Pr\{ \theta \,{<}\, {-}\FOV, \Phi \,{-}\, \theta \,{>}\, \FOV, \theta \,{<}\, 0 \} \nonumber \\
& \quad + \Pr\{ \theta \,{>}\, \FOV, \Phi \,{-}\, \theta \,{>}\, \FOV, 0 \,{<}\, \theta \,{<}\, \Phi \} \nonumber \\
& \quad + \Pr\{ \theta \,{>}\, \FOV, \Phi \,{-}\, \theta \,{<} {-}\FOV, \theta \,{>}\, \Phi \} \nonumber \\
&= \Pr\{\theta \,{<} {-}\FOV\} \,{+} \Pr\{ \FOV \,{<}\, \theta \,{<}\, \Phi \,{-}\, \FOV \} \,{+} \Pr\{\theta \,{>}\, \Phi \,{+}\, \FOV\} \nonumber \\
&= \bigtriangleup_\theta(\FOV, \Phi \,{-}\, \FOV) \,{+}\, F_\theta({-}\FOV) \,{+}\, 1 \,{-}\, F_\theta(\Phi \,{+}\, \FOV), \nonumber
\end{align}
which verifies \eqref{eqn:p4}, and completes the proof of the cdf in \eqref{eqn:cdf_nonrand_twoled_4}. \hfill\IEEEQEDhere
\end{case}  

In order to derive the desired pdf, we first observe that the cdf in \eqref{eqn:cdf_nonrand_twoled_4} is composed of the functions $\bigtriangleup_\theta(a,b)$ and $\bigtriangledown_\theta(a,b,c,d)$, given in \eqref{eqn:delta_theta} and \eqref{eqn:delta_two_int_2}, respectively, for which the possible nonzero output terms involve the cdf of the random angle $\theta$ appearing as ${-}F_\theta(a)$, $F_\theta(b)$, ${-}F_\theta(c)$, and $F_\theta(d)$. Because the partial derivative of $\bigtriangleup_\theta(a,b)$ and $\bigtriangledown_\theta(a,b,c,d)$ can be expressed as 
\begin{align}\label{eqn:delta_del}
\frac{\partial}{\partial x} \bigtriangleup_\theta(a, b) &= \begin{cases}
\frac{\partial F_\theta(b)}{\partial x} \,{-}\, \frac{\partial F_\theta(a)}{\partial x} &\mbox{for } a \leq b \\
0 & \mbox{otherwise}
\end{cases},
\end{align}
and
\begin{align}\label{eqn:delta_two_del}
\frac{\partial}{\partial x} \bigtriangledown_\theta(a, b, c, d) &= \begin{cases}
\frac{\partial F_\theta(b)}{\partial x} \,{-}\, \frac{\partial F_\theta(a)}{\partial x} &\mbox{for } c \leq a,\,  d > b \\
\frac{\partial F_\theta(d)}{\partial x} \,{-}\, \frac{\partial F_\theta(c)}{\partial x} &\mbox{for } c > a ,\, d \leq b \\
\frac{\partial F_\theta(d)}{\partial x} \,{-}\, \frac{\partial F_\theta(a)}{\partial x} &\mbox{for } c \leq a,\, d \leq b \\
\frac{\partial F_\theta(b)}{\partial x} \,{-}\, \frac{\partial F_\theta(c)}{\partial x} &\mbox{for } c > a,\,  d > b \\
0 & \mbox{otherwise}
\end{cases},
\end{align}
any individual derivative in \eqref{eqn:pdf_nonrand_twoled} can be computed using \eqref{eqn:cdf_theta_del} with proper choice of the transformation variables $u$ and $v$. Note that for constant entries which are function of only $\FOV$ and $\Phi$, we have $v\,{=}\,0$ by definition, and the partial derivative in \eqref{eqn:cdf_theta_del} yields $0$. 

As an example, the derivative of \eqref{eqn:app:p11}, which is actually the first term of the desired derivative $\frac{\partial P_1(x)}{\partial \theta}$ to be involved in the pdf expression in \eqref{eqn:pdf_nonrand_twoled}, can be given by \eqref{eqn:delta_two_del} where
\begin{align*}
a &\,{=}\, {-}\FOV \rightarrow (u,v)\,{=}\,({-}\FOV,0) \mbox{ and } \frac{\partial F_\theta(a)}{\partial x} \,{=}\, 0 \,, \\
b &\,{=}\, {-}z_1(x) \rightarrow (u,v){=}(0,{-}1) \mbox{ and } \\
& \frac{\partial F_\theta(b)}{\partial x} {=} \left[ 4x \left( c_1\,{-}\,x \right) \right]^{{-}\frac{1}{2}} f_\theta\left( {-}z_1(x) \right) \,, \\
c &\,{=}\, \Phi \,{-}\, \FOV \rightarrow (u,v)\,{=}\,(\Phi \,{-}\, \FOV,0) \mbox{ and } \frac{\partial F_\theta(c)}{\partial x} \,{=}\, 0  \,,  \\
d &\,{=}\, \Phi\,{-}\,z_2(x) \rightarrow (u,v)\,{=}\,(\Phi,{-}1) \mbox{ and } \\
& \frac{\partial F_\theta(d)}{\partial x} \,{=}\, \left[ 4x \left( c_2\,{-}\,x \right) \right]^{{-}\frac{1}{2}}f_\theta\left( \Phi\,{-}\,z_2(x) \right) \,.
\end{align*}
All the other required derivatives for \eqref{eqn:pdf_nonrand_twoled} can be computed similarly. For the situations where the argument of the cdf $F_\theta(\cdot)$ involve $\min(\cdot)$ and $\max(\cdot)$ functions, the above strategy should be applied to the argument qualifying to be the minimum or the maximum, respectively. 

As a final remark, the effective channel can only be zero when both LEDs are outside the receiver FOV, where this situation is represented by $P_4$ in the cdf expression \eqref{eqn:cdf_nonrand_twoled_4}. Because this intuition implies that $\Pr\{h^2_{\rm eff} \,{=}\,0 \} \,{=}\,P_4$, and since $\Pr\{h^2_{\rm eff} \,{<}\,0 \} \,{=}\,0$, the cdf function has a discontinuity at $x\,{=}\,0$, which appears as a Dirac delta $\delta(x)$ in the pdf expression \eqref{eqn:pdf_nonrand_twoled} with the magnitude $P_4$. \hfill\IEEEQEDhere

\bibliographystyle{IEEEtran} 
\bibliography{mypaper}
\end{document}